\documentclass[twocolumn, superscriptaddress, nobibnotes, nofootinbib, aps, prd]{revtex4-2}
\usepackage{graphicx}
\usepackage{amssymb}
\usepackage{amsmath}
\usepackage{braket}
\usepackage{multirow}
\usepackage{tikz}
\usepackage{upgreek}
\usepackage{hyperref}
\hypersetup{colorlinks=true,linkcolor=blue,filecolor=blue,citecolor = blue, urlcolor=cyan}

\usetikzlibrary{quotes, angles}

\usepackage{ulem}
\begin{document}

\title{Inertial Torsion Noise in Matter-Wave Interferometers for Gravity Experiments}

\author{Meng-Zhi Wu}
    \email{mengzhi.wu@rug.nl}
    \affiliation{Van Swinderen Institute for Particle Physics and Gravity, University of Groningen, 9747 AG Groningen, the Netherlands }

\author{Marko Toro\v{s}}
    \affiliation{Department of Physics and Astronomy, University College London, Gower Street, WC1E 6BT London, UK}
    \affiliation{Faculty of Mathematics and Physics, University of Ljubljana, Jadranska 19, SI-1000 Ljubljana, Slovenia}

 \author{Sougato Bose}
     \affiliation{Department of Physics and Astronomy, University College London, Gower Street, WC1E 6BT London, UK}

\author{Anupam Mazumdar} 
    \email{anupam.mazumdar@rug.nl}
    \affiliation{Van Swinderen Institute for Particle Physics and Gravity, University of Groningen, 9747 AG Groningen, the Netherlands }

\begin{abstract}
    Matter-wave interferometry is susceptible to non-inertial noise sources, which can induce dephasing and a resulting loss of interferometric visibility. Here, we focus on inertial torsion noise (ITN), which arises from the rotational motion of the experimental apparatus suspended by a thin wire and subject to random external torques. We provide analytical expressions for the ITN noise starting from generalized Langevin equations describing the experimental box which can then be used together with the transfer function to obtain the dephasing factor. We verify the theoretical modelling and the validity of the approximations using Monte Carlo simulations, obtaining good agreement between theory and numerics. As an application, we estimate the size of the effects for the next generation of interferometry experiments with femtogram particles, which could be used as the building block for entanglement-based tests of the quantum nature of gravity. We find that the ambient gas is a weak source of ITN, posing mild restrictions on the ambient pressure and temperature. We discuss the general ITN constraints by assuming a Langevin equation parameterized by three phenomenological parameters. 
\end{abstract}

\maketitle

\section{Introduction}\label{section 1}

Matter-wave interferometry has many salient applications for gravitational physics with devices spanning gravimeters~\cite{Peters1999}, gradiometers~\cite{science.1135459, Stray2022}, accelerometers~\cite{Saywell2023} and gyroscopes~\cite{PhysRevLett.126.197702}. 
They can also be used for fundamental physics, such as testing the equivalence principle~\cite{PhysRevLett.120.183604, PhysRevLett.125.191101, Bose:2022czr} and the quantum gravity induced entangled of masses (QGEM) to test the quantum nature of gravity in a lab~\cite{PhysRevLett.119.240401,ICTS}, see also~\cite{PhysRevLett.119.240402,Marshman:2019sne,Carney:2021vvt,Biswas:2022qto}.
The paper~\cite{Biswas:2022qto} provides the protocol to test the spin-2 nature of the graviton in an analogue of the light-bending experiment, see also~\cite{Carney:2021vvt}. One can also probe the nature of massive graviton~\cite{Elahi:2023ozf} and non-local gravitational interaction~\cite{Vinckers:2023grv} motivated by string theory.
Furthermore, some groups even consider building gravitational wave observatories based on matter-wave interferometry such as the matter-wave laser interferometer gravitation antenna (MIGA)~\cite{refId0, Junca_2019}, the matter-wave atomic gradiometer interferometric sensor (MAGIS-100)~\cite{Abe_2021, Mitchell:2022zbp}, as well as the mesoscopic interference for metric and curvature (MIMAC) scheme~\cite{Marshman_2020}.

Future matter-wave interferometry aims to exploit the regime of large masses, large superposition sizes, and long coherence times, allowing the probe of exquisitely small experimental signals. For example, the QGEM experiment would ideally require a test mass of $\sim 10^{-15}$\,kg, a superposition size of $\sim 100\,\rm\mu m$, and a coherence time of $\sim$1\,s~\cite{PhysRevLett.119.240401,Bose:2022czr,vandeKamp:2020rqh,PhysRevResearch.6.013199,Schut:2023eux}. One of the most promising setups towards this kind of experiments is the adaptation of the Stern-Gerlach interferometer (SGI) to nanoparticles~\cite{Keil2021}. SGIs based on atom chips~\cite{Machluf2013-rg} have already achieved the superposition size and coherence time of 3.93\,$\rm\mu m$ and 21.45\,ms for the half-loop configuration~\cite{Machluf2013}, respectively, and $0.38\,\rm\mu m$ and 7\,ms for the full-loop configuration~\cite{SGI_experiment}, respectively. The next generation of SGIs is currently under theoretical and numerical investigation~\cite{PhysRevLett.117.143003, PhysRevLett.125.023602, PhysRevResearch.4.023087, PhysRevResearch.4.043157, PhysRevA.107.032212}. 

An essential challenge of matter-wave interferometry is to tame the numerous noise sources, which can cause random phase fluctuations, resulting in dephasing and the loss of interferometric visibility. Vibrations of the experiment apparatus can result in residual acceleration noise (RAN)~\cite{PhysRevA.102.040202,PhysRevResearch.3.023178}, external sources of gravity can induce gradient gradient noise (GGN)~\cite{PhysRevResearch.3.023178,PhysRevD.107.104053, Junca_2019,Mitchell:2022zbp}, and charged or dipolar environmental particles can induce several electromagnetic channels of dephasing~\cite{PhysRevA.109.033301,PhysRevA.110.022412}, besides gravitational decoherence~\cite{Toros:2020krn} of the QGEM.

This paper will study the dephasing caused by the residual rotational or inertial torsion noise (ITN) for an asymmetric nanoparticle matter-wave interferometer which is sensitive to gravity-gradients~\cite{Marshman_2020,PhysRevResearch.3.023178,PhysRevD.107.104053}. ITN arises naturally in any setup whenever the experimental apparatus is subject to random torques, placing it in non-inertial rotational motion. As we will see, ITN can induce relative random phases in an interferometric experiment, resulting in a loss of interferometric contrast. Here, we will be primarily interested in understanding ITN, and we will focus on a simple single-stage suspension forming a torsion pendulum, i.e., matter-wave interferometry performed inside a hanging box. Analogous configurations of the experimental apparatus have been investigated previously in gravitational wave observatories like LIGO~\cite{harry2010advanced} and VIRGO~\cite{10.1063/1.1149783}. More advanced setups could employ additional structures like the inverted pendulum~\cite{10.1063/1.1144774} and the Roberts linkage~\cite{DUMAS20103705}.

In Sec.~\ref{section 2}, we first introduce the concept of ITN, illustrating it for matter-wave interferometry that can be modelled using qubits. In Sec.~\ref{section 3}, we generalize the analysis using linear response theory, providing the transfer function and its relation to the dephasing factor.  In Sec.~\ref{section 4}, we investigate ITN caused by ambient gas collisions on the experimental box starting from a generalized classical Langevin equation. We will compute the power spectrum density (PSD) of the ITN using the convolution theorem and verify the validity of the approximations using Monte Carlo simulations. In Sec.~\ref{section 5}, we obtain the resulting constraints on the ambient pressure and temperature and general constraints on ITN by parametrizing a generic Langevin equation modelling the experimental box with three phenomenological parameters in the dynamical equation of the experimental box. In Sec.~\ref{section 6}, we conclude with a summary of the obtained results. In Appendix \ref{appendixA}, we construct the ITN Lagrangian starting from Fermi normal coordinates and transforming to a rotating reference frame. In Appendix \ref{appendixB}, we provide for completeness and also the complete derivation of the ITN PSD using complex analysis. In Appendix \ref{appendixC}, we analyze the ITN caused by the ambient thermal gas collision.


\section{Concept of Inertial Torsion Noise and Dephasing}\label{section 2}

\begin{figure*}
    \centering
  \begin{minipage}{0.45\textwidth}
        \begin{tikzpicture}
            \draw[line width=1.25pt, inner color=white, outer color=cyan!5] (-2,0) rectangle (2,7);
            \draw[line width=2pt] (0,4)--(0,7);
            \draw[<->, orange] (-0.35,4.5) arc(150:390:0.4 and 0.15) node[above right, align=left] {torsion\\ noise};
            \draw[inner color=blue!5, outer color=blue!10] (-1,2) rectangle (1,4);

            \draw[gray, fill=blue!20] (-0.7,2.75) circle (0.1);
            \draw[gray, fill=blue!60] (-0.1,2.75) circle (0.1);
            \draw[dotted] (-0.7,2.75)--(-0.1,2.75);

            \draw[->, blue!70, line width=1.25pt] (0.8,2.2)--(1.6,2.2) node[right, black] {x};
            \draw[->, blue!70, line width=1.25pt] (0.8,2.2)--(0.8,3.4) node[above, black] {z};
        \end{tikzpicture}
     \end{minipage}
    \begin{minipage}{0.45\textwidth}
            \begin{tikzpicture}[scale=1.2]
            \draw[->, gray] (-2.5,0)--(3,0) node[right, black] {$x$};
            \draw[->, gray] (0.75,-0.5)--(0.75,4.5) node[above, black] {$t$};
        
            \draw (1.5,-0.25)--(1.5,0.25);
            \draw[thick, densely dotted] (1.5,0.25) .. controls (1.5,1) and (2,0.5) .. (2.5,1.5);
            \draw[thick, densely dotted] (2.5,1.5)--(2.5,2.25);
            \draw[thick, densely dotted] (2.5,2.25) .. controls (2,3.25) and (1.5,2.75) .. (1.5,3.5);
            \draw[thick, densely dotted] (1.5,0.25)--(1.5,3.5);
            \draw (1.5,3.5)--(1.5,4);
            
            \draw[fill=gray] (1.5,-0.25) circle (0.04);
            \draw[fill=gray] (1.5,0.25) circle (0.04);
            \draw[fill=gray] (1.5,1.7) circle (0.04);
            \draw[fill=gray] (2.5,1.7) circle (0.04);
            \draw[fill=gray] (1.5,3.5) circle (0.04);
            \draw[fill=gray] (1.5,4) circle (0.04);
        
            \draw[<->] (1.55,1.7)-- node[above]{$\Delta x$} (2.45,1.7);
            \draw[dashed] (-2.1,0.25)--(2.75,0.25);
            \draw[dashed] (-2.1,1.5)--(2.75,1.5);
            \draw[dashed] (-2.1,2.25)--(2.75,2.25);
            \draw[dashed] (-2.1,3.5)--(2.75,3.5);
            \draw[<->] (-1.9,0.3)-- node[left]{$2t_a$} (-1.9,1.45);
            \draw[<->] (-1.9,1.55)-- node[left]{$t_e$} (-1.9,2.2);
            \draw[<->] (-1.9,2.3)-- node[left]{$2t_a$} (-1.9,3.45);
        
            \node[right] at (-1.65, 0.9) {creation}; 
            \node[right] at (-1.75, 1.875) {free flight}; 
            \node[right] at (-1.8, 2.9) {recombination};       
        \end{tikzpicture}
    \end{minipage}
    \caption{(a) Experimental scheme and illustration of inertial torsion noise (ITN). The matter-wave particle (light and dark blue circles) is placed inside the experimental box (blue box), which is suspended by a thin wire (black vertical line). The box can rotate around the z-axis, while the interferometric protocol is performed along the horizontal x-axis. ITN generates random torques on the experimental box (blue rectangle), placing it in non-inertial rotational motion. The random non-inertial rotational motion induces random relative phases, which can lead to dephasing and the loss of interferometric contrast.  (b) The structure of a Stern-Gerlach interferometer with spin-1~\cite{Keil2021}. The left path keeps the same position, while the right path accelerates and decelerates at a constant rate, resulting in a superposition size $\Delta x$. Such motion is achieved using a constant magnetic field gradient, i.e. $B(x)=B_0+\eta x$, where the gradient $\eta$ is positive during $0\sim t_a$ and $3t_a+t_e\sim4t_a+t_e$, while it is negative during $t_a\sim2t_a$ and $2t_a+t_e\sim3t_a+t_e$. During the intermediate free flight of duration $t_e$, the magnetic field is uniform, and the right path does not accelerate. Such an asymmetrically shaped interferometer can be used to detect gravity-gradients~\cite{Marshman_2020,PhysRevResearch.3.023178,PhysRevD.107.104053}}
     \label{Fig1}
\end{figure*}
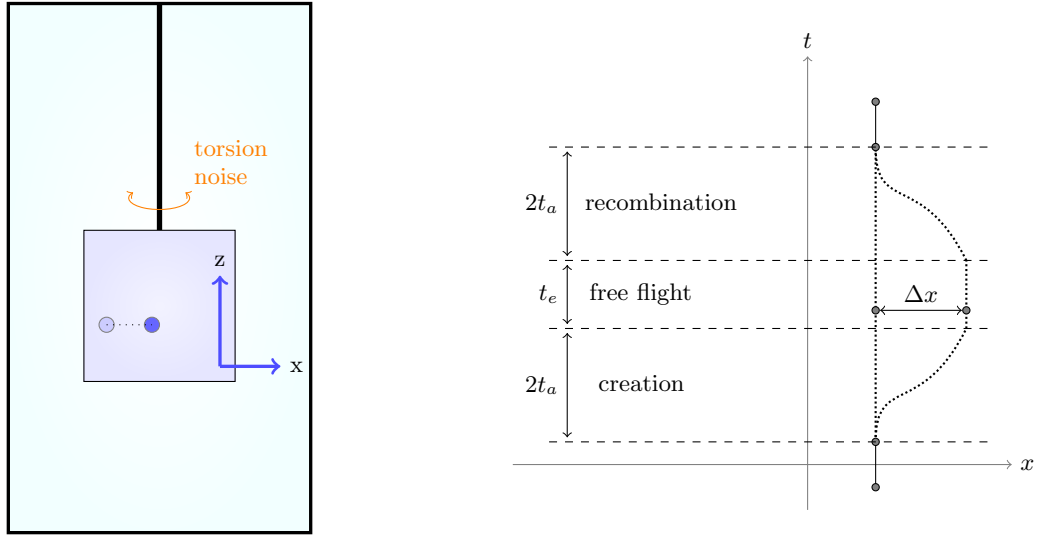

Suppose the interferometer is set up in a suspended experiment apparatus, shown as Fig.~\ref{Fig1}(a). A natural reference frame is the comoving frame of the experimental apparatus. 
For instance, consider a Stern-Gerlach interferometer controlled by a static magnetic field $\mathbf{B}(x,y,z)$, which holds for the comoving frame of the experiment apparatus. If the SGI is studied under another reference frame which moves relative to the comoving reference, then the Lorentzian transformation of the magnetic field $\mathbf{B}(t)$ has to be taken into account, and even the induced electric field also needs to be studied. The comoving reference frame of the experiment apparatus is thus preferred as it simplifies the analysis.

However, the experimental apparatus itself can also be shaken by various environmental disturbances, such as vibrations of mechanical supports and jitter caused by collisions of air molecules. These vibrations of the experimental apparatus are applied to the interferometer as non-inertial forces, resulting in acceleration and rotation noises, affecting the interferometer's final phase, i.e., \emph{residual acceleration noise} (RAN) \cite{PhysRevResearch.3.023178, PhysRevD.107.104053} and \emph{inertial rotation noise} (ITN) which we investigate in this paper. Such noises can induce dephasing, which can be mitigated only by carefully controlling the experimental setup and its environment. 

In this paper, we will focus on the Stern-Gerlach interferometer with the configuration shown in Fig.~\ref{Fig1}(b), which can be achieved by a system with a spin-1 state and spin-0 state (embedded in an object) in a magnetic field with a constant gradient, see \cite{Marshman_2020}. In particular, the system is prepared in a spin superposition of $\ket{S=0}$ and $\ket{S=1}$, say $\ket{\psi}=1/\sqrt{2}(\ket{S=0}+\ket{S=1})$. At the same time, the external magnetic field has a linear spatial distribution in the $x$-axis, i.e. $B(x)=B_0+\eta x$, where the magnetic gradient $\eta$ will flip several times to accelerate and decelerate the state $\ket{S=1}$. 

Therefore, the acceleration of the $\ket{S=1}$ state is a constant $a_m=g\mu_B\eta$ during $0\sim t_a$ and $3t_a+t_e\sim4t_a+t_e$, and $a_m=-g\mu_B\eta$ during $t_a\sim2t_a$ and $2t_a+t_e\sim3t_a+t_e$, where $g=2$ is the Lande factor and $\mu_B=9.27\times10^{-24}$\,J/T is the Bohr magneton. 

In an ideal experiment (without any dephasing and decoherence), the final state of the system is
\begin{equation}
    \ket{\psi} = \frac{1}{\sqrt{2}}\mathrm{e}^{\mathrm{i}\phi_{\rm global}}(\ket{S=0,L}+\mathrm{e}^{\mathrm{i}\phi_{\rm diff}}\ket{S=1,R}),
\end{equation}
where $\phi_{\rm global}$ is the global phase of the quantum state and $\phi_{\rm diff}$ is the differential phase between two paths denoted as "L" and "R". The global phase does not have observable effects, while the differential phase $\phi_{\rm diff}$ usually encodes the signal we want to extract. For example, $\phi_{\rm diff}$ is related to the gravitation acceleration $g$ for a gravimeter~\cite{PhysRevLett.111.180403}. Another example is the QGEM experiment, which encodes information about the nature of gravity. 

However, some classical noises like RAN and ITN always exist. They will contribute some random phase $\delta\phi$ on $\phi_{\rm diff}$.
Although $\mathrm{e}^{\mathrm{i}\delta\phi}$ itself is a pure phase, the ensemble average of such a random phase can lead to a damping factor, known as a dephasing effect. For ease of convenience, $\delta\phi$ is supposed to follow a Gaussian distribution with a mean value of zero and a variance of $\Gamma\equiv\mathbb{E}[(\delta\phi)^2]$/2, then the expectation value of the phase can be computed as
\begin{equation}
    \mathbb{E}[\mathrm{e}^{\mathrm{i}\delta\phi}] = \mathrm{e}^{-\mathbb{E}[(\delta\phi)^2]} \equiv \mathrm{e}^{-\Gamma},
\end{equation}
where $\mathbb{E}[\cdot]$ is the ensemble average of random variables. Note that for a more general probability distribution of $\delta\phi$, one can also define a similar $\Gamma$ to describe the dephasing effect~\cite{Lombardo_2006}. Such decay factor $\mathrm{e}^{-\Gamma}$ will cause the loss of visibility of the interferometer. In particular, consider the ensemble average of the density matrix $\hat{\rho}=\ket{\psi}\bra{\psi}$:
\begin{equation}\label{density matrix}
\begin{split}
    \mathbb{E}[\hat{\rho}] &= \frac{1}{2}\bigg( \ket{L}\bra{L} + \ket{R}\bra{R} \\
        &+ \mathbb{E}[\mathrm{e}^{\mathrm{i}\delta\phi}]\mathrm{e}^{2\mathrm{i}\phi_{\rm diff}}\ket{R}\bra{L} + \mathbb{E}[\mathrm{e}^{-\mathrm{i}\delta\phi}]\mathrm{e}^{-2\mathrm{i}\phi_{\rm diff}}\ket{L}\bra{R} \bigg),
\end{split}
\end{equation}
the off-diagonal terms decay exponentially with the damping given by the variance $\Gamma$. Consequently, the expectation value $\text{Tr}(\hat{W}\hat{\rho})$ of any witness operator $\hat{W}$ will also decay with respect to the factor $\mathrm{e}^{-\Gamma}$. 

For example, one can consider two applications of matter-wave interferometers for gravity experiments.
\begin{itemize}
    \item The first application is the gravimeter based on a NV center~\cite{PhysRevLett.111.180403, PhysRevA.88.033614}. The gravitational acceleration $g$ is proportional to the differential phase as $\phi_{\rm diff}=16\pi mg\Delta z/(\hbar\omega_0)$, where $\Delta z$ is the superposition size along the z-axis, and $\omega_0\sim100$\,kHz is the trapping frequency. 
    Then the phase fluctuation can cause a sensitivity loss on the measurement result of $g$, then one can obtain a constrain for $\Gamma$ by the variance $\sigma_g$ of $\delta g$ as $\sqrt{\Gamma}=16\pi m\sigma_g\Delta z/(\hbar\omega_0)$, as long as $\delta\phi_{\rm diff}$ and $\delta g$ follow Gaussian distribution. Choosing the values as $m\sim10^{-16}$\,kg, $\Delta z\sim10^{-8}$\,m~\cite{Chen:18}, then one may obtain a threshold as $\sqrt{\Gamma}<\sigma_g10^7\,\rm{s^2/m}$. A common performance of gravimeters is $\sigma_g=10^{-9}\sim10^{-10}g$~\cite{Chen:18}, then one can choose the threshold of $\Gamma$ as $10^{-6}$.
    \item Another example is the QGEM experiment~\cite{PhysRevA.110.022412, PhysRevResearch.3.023178}, which uses two interferometers coupled by gravity to investigate gravity-induced quantum entanglement. The witness $\hat{W}$ is proposed as the positive partial transpose (PPT) witness, which, in the case of two qubits, provides a sufficient and necessary condition for entanglement based on the Peres-Horodecki criterion~\cite{RevModPhys.81.865, PhysRevA.105.032411, PhysRevA.104.052416}, which requires the expectation value of the witness satisfying $\braket{\hat{W}}<0$.
    As is calculated in \cite{PhysRevA.110.022412}, the expectation value of the witness under the dephasing effect is $\braket{\hat{W}}=(1-\mathrm{e}^{-2\Gamma})/4-\mathrm{e}^{-\Gamma/2}\sin\phi_g$, then the constrain on $\Gamma$ is that $\Gamma/2<\mathrm{e}^{-\Gamma/2}\sin\phi_g\approx\phi_g$. The value of $\phi_g$ has been estimated as $\phi_g\approx0.015$~\cite{PhysRevA.102.062807, PhysRevResearch.3.023178}, then we will choose a threshold $\Gamma=0.01$ for further discussions on the values of parameters in this paper.
\end{itemize}

Dephasing, together with all other types of decoherence, can be also characterized by computing the purity:
\begin{equation}\label{purity}
    \text{Tr}(\mathbb{E}[\hat{\rho}]^2) = \frac{1}{2}(1+\mathrm{e}^{-2\Gamma}) \approx 1 - \Gamma.
\end{equation}
Note that the dephasing comes from the ensemble average of the density matrix. In particular, $\text{Tr}(\hat{\rho}^2)$ without ensemble average still equals $1$. We will refer to $\Gamma$ as the \emph{dephasing factor} in the following text. In the next section, we will analyze the dephasing factor $\Gamma$ caused by the ITN.

\section{Dephasing as Linear Response to Inertial Torsion Noise}\label{section 3}

As is derived in Appendix \ref{appendixA}, the Lagrangian of the inertial torsion noise is given by
\begin{equation}\label{Lagrangian}
    L_{\rm ITN} = \frac{1}{2}m\dot{\Theta}^2x^2,
\end{equation}
where $m$ and $x$ are the interferometer's mass and position, and $\Theta(t)$ is the torsion angle of the suspended apparatus, i.e., the angle between the apparatus reference frame and the inertial reference frame. 

The angle $\Theta(t)$ is assumed to be a stochastic Gaussian process, satisfying the following two properties:
\begin{equation}
\begin{aligned}
    \mathbb{E}[\Theta(t)] &= 0 \\
    \mathbb{E}[\Theta(t_1)\Theta(t_2)] &= \int S_{\Theta\Theta}(\omega)\mathrm{e}^{-\mathrm{i}\omega(t_2-t_1)}\mathrm{d}\omega
\end{aligned}
\end{equation}
where $S_{\Theta\Theta}(\omega)$ is the \emph{power spectrum density} (PSD) of $\Theta(t)$ and the second identity is known as the Wiener-Khinchin theorem. The PSD of $\Theta$ can be measured in the experiment. In this paper, we will consider a specific source of fluctuation of $\Theta$ (e.g., the collision by gas molecules, which will discussed in the next section).

As is proved in~\cite{Storey1994TheFP, Wu:2024tcr}, the phase fluctuations due to noises are determined by the path integral of the corresponding Lagrangian of the noise along the unperturbed classical trajectories. Thus, the phase fluctuation is
\begin{equation}
\begin{split}
    \delta\phi &= \frac{1}{\hbar}\int L_{\rm ITN}[x_R(t)]-L_{\rm ITN}[x_L(t)] \mathrm{d}t \\
        &= \frac{m}{2\hbar}\int \dot{\Theta}^2(t)\left(x_R^2(t)-x_L^2(t)\right) \mathrm{d}t,
\end{split}
\end{equation}
where $x_R(t)$ and $x_L(t)$ are the trajectories of the interferometer's two arms. Assuming the expectation value of $\delta\phi$ vanishes, and the variance $\Gamma\equiv\mathbb{E}[(\delta\phi)^2]$ of the random phase can be regarded as the linear response of the interferometer to the torsion noise~\cite{PhysRevResearch.3.023178}:
\begin{equation}\label{Gamma}
    \Gamma = \frac{m^2}{4\hbar^2}\int S_{\dot{\Theta}^2\dot{\Theta}^2}(\omega)F(\omega)\mathrm{d}\omega,
\end{equation}
where $S_{\dot{\Theta}^2\dot{\Theta}^2}(\omega)$ is the PSD of the ITN. According to the Wiener-Khinchin theorem, the PSD is the Fourier transform of the autocorrelation function of the torsion noise
\begin{equation}
    S_{\dot{\Theta}^2\dot{\Theta}^2}(\omega) = \int\mathrm{E}[\dot{\Theta}^2(t_0)\dot{\Theta}^2(t_0+\tau)]\mathrm{e}^{\mathrm{i}\omega\tau}\mathrm{d}\tau.
\end{equation}
Note that $S_{\dot{\Theta}^2\dot{\Theta}^2}(\omega)$ has a unit of $\rm Hz^4/Hz$, where $\rm Hz^4$ comes from the square of $\dot{\Theta}^2$ and the denominator $\rm Hz$ describes the density of frequency space. The $F(\omega)$ in Eq.~\eqref{Gamma} has the unit $\rm m^4s^2$, and is given by
\begin{equation}\label{F definition}
    F(\omega) = \left|\int \left(x_R^2(t)-x_L^2(t)\right)\mathrm{e}^{\mathrm{i}\omega t}\mathrm{d}t\right|^2.
\end{equation}
Eq.~\eqref{Gamma} describes the input-output relation of the interferometer and only depends on the trajectories of the two arms, so it can be called the \emph{transfer function} of the interferometer\cite{PhysRevD.107.104053}. For the interferometer in Fig.\ref{Fig1}(b), the left arm is static $x_L(t)\equiv0$, and the right arm $x_R(t)$ is described by a piecewise function consisting of several quadratic functions of $t$ because the acceleration is $\pm a_m$ for the different time range. Then the transfer function can be computed as 
\begin{equation}\label{F}
\begin{split}
    F(\omega) &= 16\frac{a_m^4}{\omega^{10}}\bigg[6\omega t_a\cos\left(\omega(t_a+\frac{t_e}{2})\right) + \left(\omega^2t_a^2+3\right) \sin\frac{\omega t_e}{2} \\
        &- 3\sin\left(\omega(2t_a+\frac{t_e}{2})\right) - \omega^2t_a^2\sin\left(\omega(t_a+\frac{t_e}{2})\right) \bigg]^2.
\end{split}
\end{equation}
\begin{figure}
    \centering
    \includegraphics[scale=0.5]{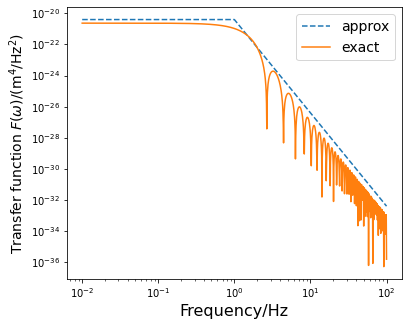}
    \caption{ Transfer function of the interferometer illustrated in Fig.\ref{Fig1}(b) as a function of the frequency $f=\omega/(2\pi)$. The time parameters are $t_a=0.25$\,s and $t_e=0$\,s. The magnetic field gradient $\eta$ is chosen as $10^4$\,T/m and the mass of the interferometer is chosen as $10^{-15}$\,kg. The resulting acceleration $a_m=1.8\times10^{-4}\,\rm m/s^2$ produces the maximum superposition size $\Delta x=a_mt_a^2=11.2\,\rm\mu m$. As is shown, the transfer function tends to a constant $C\sim(\Delta x)^4T^2\sim10^{-20}\,\rm m^4s^2$ at low frequencies, and decreases as $\omega^{-6}$ at high frequencies. The approximate transfer function from Eq.~\eqref{F approx}  (dashed blue line) captures the behaviour of the transfer function from Eq.~\eqref{F} (orange line). }
    \label{transfer function}
\end{figure}
Fig.\ref{transfer function} shows the transfer function with the parameters chosen as $t_a=0.25$\,s, $t_e=0$\,s and $a_m=1.8\times10^{-4}\,\rm m/s^2$, where $a_m=g\mu_B\eta/m_0$ with the magnetic field gradient $\eta=10^4$\,T/m and the mass of the interferometer $m_0=10^{-15}$\,kg.

As is shown in Fig.\ref{transfer function}, $F(\omega)$ tends to a constant $C\sim(\Delta x)^4T^2\sim10^{-20}\,\rm m^4s^2$ in low frequency limit
where $\Delta x=a_mt_a^2=11.2\,\rm\mu m$ and $T=4t_a+t_e$ are the superposition size and the total experiment time. It is because the factor $\mathrm{e}^{i\omega t}\to1$ in the low-frequency limit $\omega\to0$, and the superposition size $\Delta x$ is a upper bound for $x_R(t)$ when $x_L(t)\equiv0$, so $F(\omega)$ in \eqref{F definition} can be approximated in low-frequency region as
\begin{equation}
    F(\omega) \sim \left|\int(\Delta x)^2\mathrm{d}t\right|^2 = (\Delta x)^4T^2.
\end{equation}

At the high frequency limit, $F(\omega)$ decrease as $\omega^{-6}$. In particular, the trajectory $x_R(t)$ can be generally expanded as a Taylor series of $t$ with the leading order $x_R(t)\sim t$, so $x^2_R(t)\sim t^2$, then the leading order of $F(\omega)$ can be estimated according to \eqref{F definition} as
\begin{equation}
    F(\omega) \sim \left|\int t^2\mathrm{e}^{i\omega t}\mathrm{d}t\right|^2 \sim (\omega^{-3})^2 = \omega^{-6}.
\end{equation}

Combining the low-frequency and high-frequency behaviours of $F(\omega)$, one can use the Heaviside step function $\theta(\cdot)$ to approximately describe $F(\omega)$ as 
\begin{equation}\label{F approx}
    F(\omega) \approx (\Delta x)^4T^2\left(\theta\left(\frac{2\pi}{T}-\omega\right) + \left(\frac{2\pi}{T\omega}\right)^6\theta\left(\omega-\frac{2\pi}{T}\right)\right).
\end{equation}

\section{Power Spectrum Density of Inertial Torsion Noise}\label{section 4}

In this section, we will analyze the inertial torsion noise caused by the thermal motion of gas molecules surrounding the experimental box.

According to the convolution theorem, the PSD of ITN is the self-convolution of the PSD of the random motion of the suspended apparatus
\begin{equation}\label{ITN convolution}
    S_{\dot{\Theta}^2\dot{\Theta}^2}(\omega) = S_{\dot{\Theta}\dot{\Theta}}(\omega)\ast S_{\dot{\Theta}\dot{\Theta}}(\omega),
\end{equation}
where $S_{\dot{\Theta}\dot{\Theta}}(\omega)$ is the PSD of $\dot{\Theta}$. In frequency space, there is a correspondence that $\dot{\Theta}\sim\mathrm{i}\omega\Theta$, so the PSD of $\dot{\Theta}(t)$ is
\begin{equation}\label{PSD of dot Theta}
    S_{\dot{\Theta}\dot{\Theta}}(\omega) = \omega^2 S_{\Theta\Theta}(\omega). 
\end{equation}
As is shown in Fig.\,\ref{Fig1}(a), the rotational motion of the experimental box can be modelled as a torsion pendulum, which follows a generalized Langevin equation
\begin{equation}\label{generic Langevin}
    \ddot{\Theta} = -\Omega_{\rm rot}^2\Theta - \gamma\dot{\Theta} + \sqrt{A}\Theta_{\rm in}(t).
\end{equation}
The input random noise term $\Theta_{\rm in}(t)$ is a unit delta-correlated stationary Gaussian process with zero-mean, satisfying $\mathbb{E}[\Theta_{\rm in}(t)] = 0$ at any time $t$, and $\mathbb{E}[\Theta_{\rm in}(t_1)\Theta_{\rm in}(t_2)] = \delta(t_1-t_2)$. The amplitude of the external noise $\sqrt{A}$~\footnote{Note that $A$ is of the unit $\rm Hz^4/Hz$. It is because the term $\sqrt{A}\Theta_{\rm in}(t)$ has a dimension $\rm [T^{-2}]$, where $\Theta_{\rm in}(t)$ has a dimension $\rm [T^{-1/2}]$.} and the dissipation rate $\gamma$ are determined by the experiment. The intrinsic torsion frequency is given by~\cite{MOHAZZABI2001677} 
\begin{equation}\label{Omega_rot}
    \Omega_{\rm rot} = \sqrt{\frac{\kappa}{I}} = \sqrt{\frac{\pi\mathbb{G}d^4}{32l I}},
\end{equation}
where $I$ is the moment of inertia of the experiment box and $\kappa=\sqrt{\pi\mathbb{G}d^4/32l}$ is the torsion constant. $\mathbb{G}$ is the shear modulus of the material of the suspension wire, $d$ and $l$ are the diameter and the length of the wire. 
The size and the mass of the experiment box can be built up as $10^{-1}\sim10^0$\,m and $10^1\sim10^2$\,kg, so the moment of inertia $I$ can be estimated as $10^0\sim10^2\,\rm kg\cdot m^2$. The parameters of the suspension wire are around $d=10^{-3}\sim10^{-2}$\,m, $l=10^0\sim10^1$\,m and $\mathbb{G}\sim10^{10}$\,Pa, then the intrinsic frequency $\Omega_{\rm rot}$ of the torsion pendulum is around $10^{-2}\sim10^1$\,Hz. 

For example, if the experiment box is built up with size $L=0.6$\,m and mass $M=30$\,kg, then the moment of inertia is $I=ML^2/6=1.8\,\rm kg\cdot m^2$. If the suspension wire is set as $d=5\times10^{-3}$\,m and $l=5$\,m, and the shear modulus is chosen as $\mathbb{G}=7.93\times10^{10}$\,Pa for steel~\cite{Crandall1972AnIT}, then the intrinsic torsion frequency is $\Omega_{\rm rot}\approx0.735$\,Hz according to Eq.~\eqref{Omega_rot}.

Based on the dynamical equation \eqref{generic Langevin} of the experiment box, the power spectrum for $\Theta$ is~\footnote{Here we used the result $S_{\rm in}(\omega)\equiv|\Theta_{\rm in}(\omega)|^2/T_{\rm tot}=1$, which can be obtained through the auto-correction condition $\mathbb{E}[\Theta_{\rm in}(t_1)\Theta_{\rm in}(t_2)] = \delta(t_1-t_2)$ and the Wiener-Khinchin theorem that
\begin{equation*}
\begin{split}
    \frac{|\Theta_{\rm in}(\omega)|^2}{T_{\rm tot}} &= \int\mathbb{E}[\Theta_{\rm in}(t)\Theta_{\rm in}(t+\tau)]\mathrm{e}^{\mathrm{i}\omega\tau}\mathrm{d}\tau \\
        &=\int\delta(\tau)\mathrm{e}^{\mathrm{i}\omega\tau}\mathrm{d}\tau = 1.
\end{split}
\end{equation*}}
\begin{equation}\label{Theta PSD}
    S_{\Theta\Theta}(\omega) = \frac{A}{(\Omega_{\rm rot}^2-\omega^2)^2+\gamma^2\omega^2}.
\end{equation}

\begin{figure}
    \centering
    \includegraphics[scale=0.6]{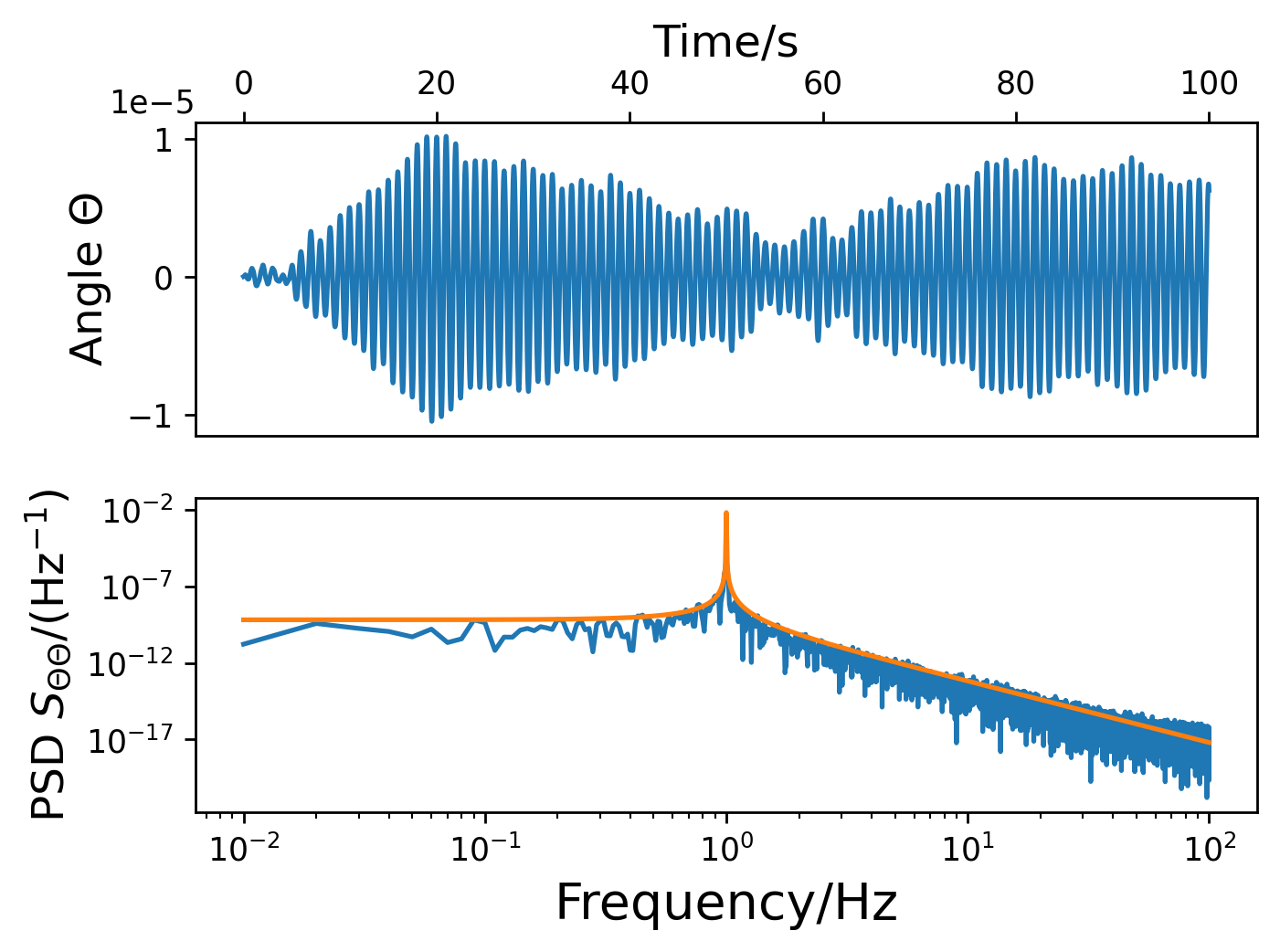}
    \caption{Monte Carlo simulation for the motion of the box. (a) Simulated time-trace of Eq.~\eqref{rotation Langevin}. (b)  The PSD $S_{\Theta\Theta}(\omega)$ in the lower plot is computed by FFT. Theoretically the PSD is a Lorentzian distribution which fits the simulation result well. The intrinsic frequency $\Omega_{\rm rot}$ is chosen as $2\pi\times1$\,Hz, the damping rate $\gamma$ is chosen as $10^{-10}$\,Hz and the noise amplitude $\sqrt{A}$ is chosen as $10^{-3}\,\rm Hz^2/\sqrt{Hz}$}.
    \label{Langevin equation MC}
\end{figure}

One may do a Monte Carlo simulation of the Langevin equation given in Eq.~\eqref{generic Langevin}, shown as the upper figure of Fig.~\ref{Langevin equation MC}, and calculate the corresponding PSD by fast Fourier transform (FFT), where a Hanning window has been added to avoid the spectral leakage, shown as the lower figure of Fig.~\ref{Langevin equation MC}. We find that the analytic result Eq.~\eqref{Theta PSD} matches well with the numerical result.

Since the PSD of ITN is the self-convolution of $S_{\dot{\Theta}\dot{\Theta}}(\omega)$ according to Eq.~\eqref{ITN convolution}, one can compute the PSD of the ITN driven by the external noise with the PSD given by Eq.~\eqref{Theta PSD} as
\begin{equation} \label{ITN PSD}
    S_{\dot{\Theta}^2\dot{\Theta}^2}(\omega) = A^2\frac{4\omega^4+4(\gamma^2-3\Omega_{\rm rot}^2)\omega^2+16\Omega_{\rm rot}^4}{\gamma(\omega^2+\gamma^2)(4\gamma^2\omega^2+(\omega^2-4\Omega_{\rm rot}^2)^2)}. 
\end{equation}
\begin{figure}
    \centering
    \includegraphics[scale=0.5]{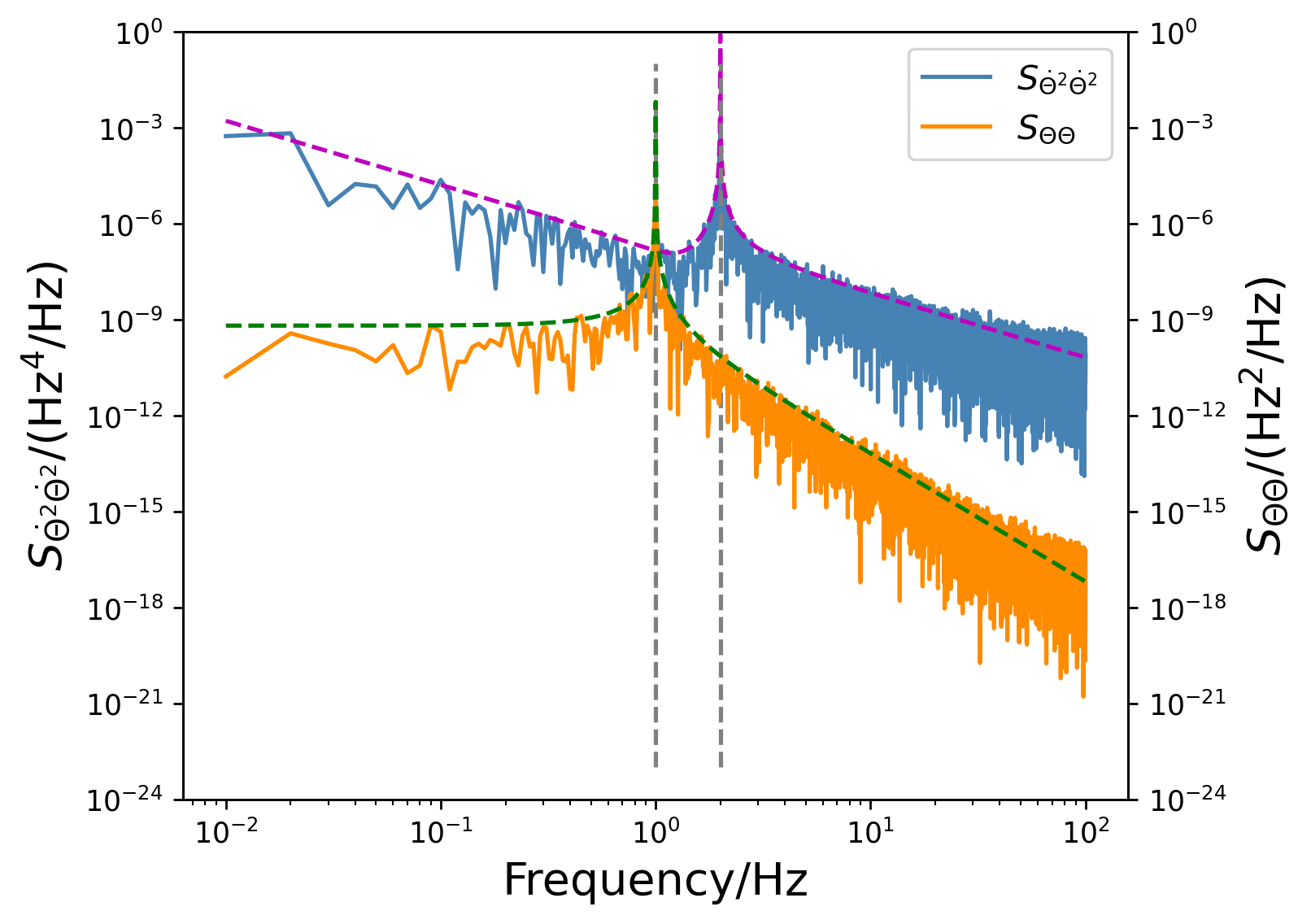}
    \caption{PSD of ITN $S_{\dot{\Theta}^2\dot{\Theta}^2}(\omega)$ and angular motion PSD $S_{\Theta\Theta}(\omega)$. The parameters are chosen as $\Omega_{\rm rot}=2\pi\times1$\,Hz, $\gamma=10^{-10}$\,Hz, I=10\,$\rm kg\cdot m^2$ and $\sqrt{A}=10^{-3}\,\rm Hz^2/\sqrt{Hz}$. . The peak frequency of $S_{\dot{\Theta}^2\dot{\Theta}^2}(\omega)$ is doubled compared to $S_{\Theta\Theta}(\omega)$ as expected when computing the square of the noise~\footref{footnoteexp1}. The dashed green and purple lines show that the theoretical expressions in Eqs.~\eqref{Theta PSD} and ~\eqref{ITN PSD} fit well the simulated PSDs.}
    \label{PSD ITN}
\end{figure}
The detailed mathematical steps are summarized in Appendix \ref{appendixB}. Fig.~\ref{PSD ITN} shows the analytical result and the numerical simulation of both $S_{\Theta\Theta}(\omega)$ and $S_{\dot{\Theta}^2\dot{\Theta}^2}(\omega)$. 

The asymptotic behaviour of the PSD of ITN is different from the $S_{\Theta\Theta}(\omega)$. In low frequency limit, $S_{\dot{\Theta}^2\dot{\Theta}^2}(\omega)$ behaves like $1/(\omega^2+\gamma_{\rm rot}^2)$ and tends to a constant $A^2/\gamma$. In high frequency region, $S_{\dot{\Theta}^2\dot{\Theta}^2}(\omega)$ decreases as $\omega^{-2}$ according to the analytic result \eqref{ITN PSD}. 

There is also a peak of the PSD of ITN, of which the resonance frequency translates from $\Omega_{\rm rot}$ to $2\Omega_{\rm rot}$, which is a common property for a squared noise
\footnote{This property can be understood as follows. Since $S_{\dot{\Theta}\dot{\Theta}}$ has a peak at $\Omega_{\rm rot}$, then it can be written as a $\delta$-function plus some small function $o(\omega)$:
$$S_{\dot{\Theta}\dot{\Theta}}(\omega) = S_0\delta(\omega-\Omega_{\rm rot}) + o(\omega),$$
where $S_0$ is the amplitude of the peak. Then the self-convolution of $S_{\dot{\Theta}\dot{\Theta}}$ gives 
\begin{equation*}
\begin{split}
    S_{\dot{\Theta}^2\dot{\Theta}^2}(\omega) &= S_0^2\int\delta(\omega'-\Omega_{\rm rot})\delta(\omega-\omega'-\Omega_{\rm rot})\mathrm{d}\omega' + o'(\omega) \\
        &= S_0^2\delta(\omega-2\Omega_{\rm rot}) + o'(\omega),
\end{split}
\end{equation*}
which means the peak position locates at $2\Omega_{\rm rot}$. In Appendix \ref{appendixB}, we offer another method to understand this property in the time domain.\label{footnoteexp1}}. 
In a small damping limit $\gamma_{\rm rot}\ll\Omega_{\rm rot}$, the peak value of $S_{\dot{\Theta}^2\dot{\Theta}^2}(\omega)$ is
\begin{equation}\label{peak height}
    S_{\dot{\Theta}^2\dot{\Theta}^2}(2\Omega_{\rm rot}) \approx \frac{A^2}{2\gamma^3}.
\end{equation}
It is notatble that
\begin{equation}
    S_{\dot{\Theta}^2\dot{\Theta}^2}(2\Omega_{\rm rot}\pm\gamma) \approx \frac{A^2}{4\gamma^3}.
\end{equation}
So the frequency at which the PSD equals to the half peak value are $\omega=2\Omega_{\rm rot}\pm\gamma$ and the corresponding full width at half maximum (FWHM) is $2\gamma$. The radio between the bandwidth and the peak value is known as the quality factor (Q-factor) of the PSD, which characterizes the sharpness of the peak. Then the Q-factor of $S_{\dot{\Theta}^2\dot{\Theta}^2}(\omega)$ is
\begin{equation}
    Q_{\rm ITN} = \frac{2\Omega_{\rm rot}}{2\gamma} = \frac{\Omega_{\rm rot}}{\gamma},
\end{equation}
which is exactly the same as the Q-factor of $S_{\Theta\Theta}(\omega)$.

\section{Dephasing Factor and Experiment Parameters Constrain}\label{section 5}

Based on the result of the transfer function in Eq.~\eqref{F} or Eq.~\eqref{F approx}, and the power spectrum density \eqref{ITN PSD} of the inertial torsion noise, the dephasing factor $\Gamma$ can be computed through the integral in Eq.~\eqref{Gamma}. 
Note that the analytic solution to the integral is very complicated, so we compute it numerically. It is noteworthy that a resolution of $\omega_{\rm min}=2\pi/T_{\rm tot}$ exists as a cutoff in frequency space for the numerical calculation. Physically, this cutoff indicates that a low-frequency signal or noise will not be measurable by the experiment if the total experiment time is shorter than a single period of such signal or noise. 
\begin{figure}
    \centering
    \includegraphics[scale=0.5]{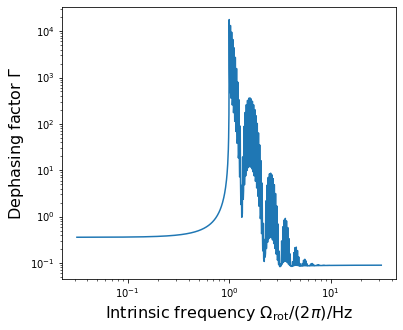}
    \caption{\small Dephasing factor $\Gamma$ as a function of the intrinsic torsion frequency $\Omega_{\rm rot}$. The parameters are chosen as $m_0=10^{-15}$\,kg, $\eta=10^4$\,T/m, $t_a=0.25$\,s and $t_e=0$\,s for the interferometer, and $\gamma=10^{-10}$\,Hz, $I=10\,\rm kg\cdot m^2$ and $A=10^{-10}\,\rm Hz^4/Hz$ for the PSD of ITN. There is a resonance between the ITN and the transfer function with the first dominant peak at $\Omega_{\rm rot}\sim2\pi/T_{\rm tot}\sim2\pi\times1$\,Hz, and additional smaller peaks visible in the range upto 10. In the low frequency limit and high frequency limit, the dephasing factor $\Gamma$ is approximately independent of the intrinsic frequency $\Omega_{\rm rot}$. }
    \label{Gamma-Omega0}
\end{figure}

Fig.~\ref{Gamma-Omega0} shows the dependence of $\Gamma$ on the intrinsic torsion frequency $\Omega_{\rm rot}$ of the experiment apparatus, where the parameters of $F(\omega)$ are chosen as $m_0=10^{-15}$\,kg, $t_a=0.25$\,s and $t_e=0$\,s, and the parameters of $S_{\dot{\Theta}^2\dot{\Theta}^2}(\omega)$ are $A=10^{-10}\,\rm Hz^4/Hz$ and $\gamma=10^{-10}$\,Hz. As is shown, there is a resonance between the ITN and the transfer function near $\Omega_{\rm rot}\sim2\pi/T_{\rm tot}=2\pi\times1$\,Hz and some harmonic resonances. Note that near the resonance peak, the precision of the numerical calculation is limited because the value is highly related to the sample points in frequency space.

When $\Omega_{\rm rot}$ is much larger than $2\pi\times10^0$\,Hz, $\Gamma$ becomes approximately independent on $\Omega_{\rm rot}$. On the other hand, when $\Omega_{\rm rot}$ is much smaller than $2\pi\times10^{-1}$\,Hz, the peak will be below the frequency cut-off $\omega_{\rm min}$, so the dephasing factor will be tiny. For an actual experiment, the low intrinsic frequency limit is expected, so the parameters $d$, $l$, $\mathbb{G}$ and $I$ in \eqref{Omega_rot} have to be carefully designed to ensure $\Omega_{\rm rot}$ much smaller than the resonance frequency $2\pi/T$ of the interferometer.
\begin{figure}
    \centering
    \includegraphics[scale=0.5]{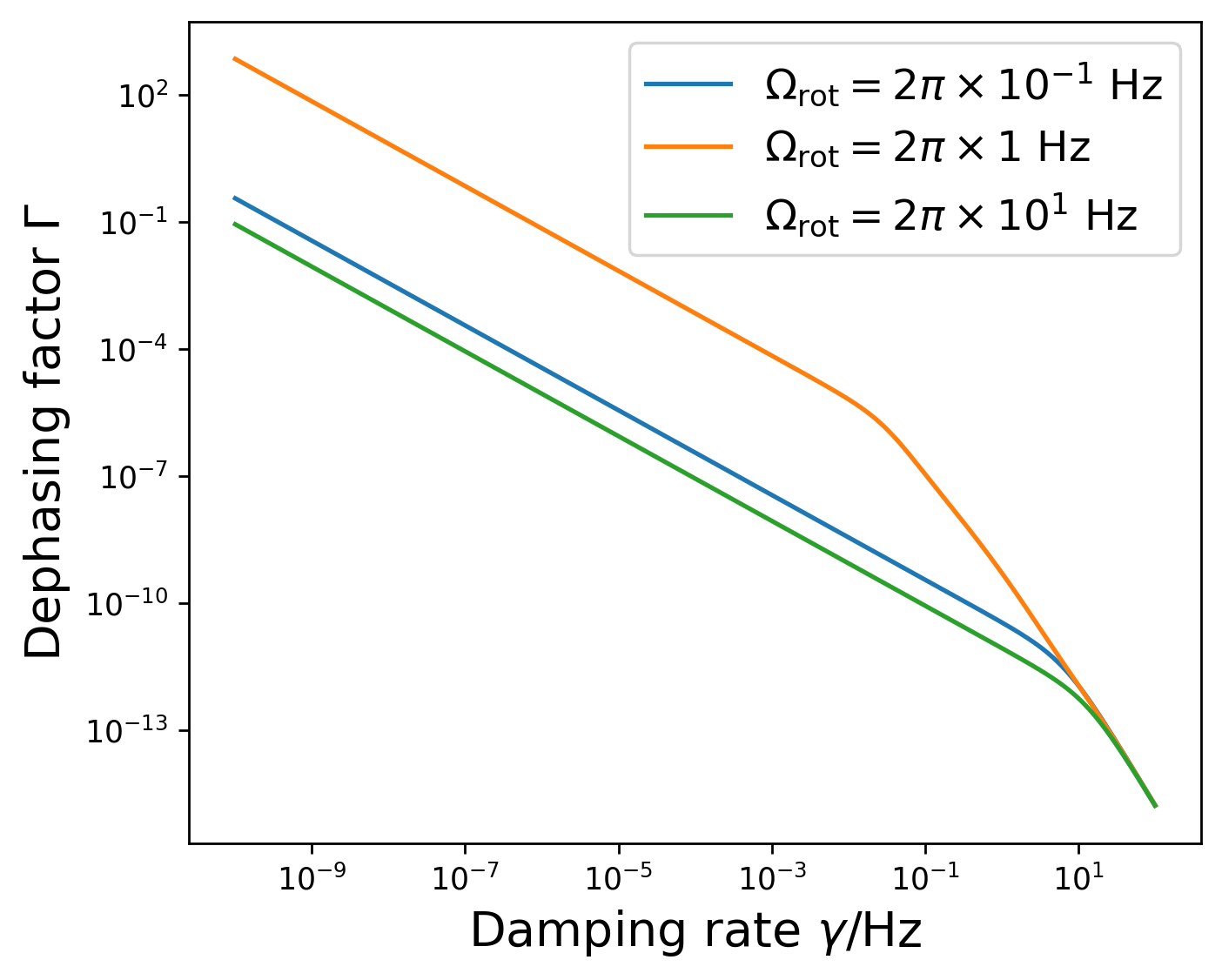}
    \caption{\small The relationship between $\Gamma$ and the damping factor $\gamma$ for a fixed amplitude $A=10^{-10}\,\rm Hz^3$. The transfer function is given in Eq.~\eqref{F} with parameters $t_a=0.25$\,s and $t_e=0$\,s. Three different $\Omega_{\rm rot}$ are chosen as the low rotation frequency limit, the high rotation frequency limit and the resonance frequency condition $\Omega_{\rm rot}\sim2\pi/T=2\pi$\,Hz. $\Gamma$ is proportional to $\gamma^{-1}$ when $\gamma\ll\Omega_{\rm rot}$, while it decreases as a speed of $\gamma^{-3}$ in the overdamped region.}
    \label{Gamma-gamma_fixA}
\end{figure}

Fig.~\ref{Gamma-gamma_fixA} shows the dependence of $\Gamma$ on the damping rate $\gamma$. As is shown, when the damping rate is very low, i.e. $\gamma\ll\Omega_{\rm rot}$, the dephasing parameter $\Gamma$ is approximately proportional to $\gamma^{-1}$ due to the $\gamma^{-1}$ factor in \eqref{ITN PSD}.

However, when $\gamma$ increases after a critical point $\gamma\sim\Omega_{\rm rot}$, the dephasing $\Gamma$ is approximately $\gamma^{-3}$. This is because the PSD of ITN at the high damping rate limit $\gamma\gg\Omega_{\rm rot}$ is approximately
\begin{equation}
    S_{\dot{\Theta}^2\dot{\Theta}^2}(\omega) \approx A^2 \frac{4\omega^4+4\gamma_{\rm rot}^2\omega^2}{\gamma(\omega^2+\gamma^2)(\omega^4+4\gamma^2\omega^2)} \sim \frac{A^2}{\gamma^3}.
\end{equation}
Physically it can be interpreted as an overdamped oscillator. In particular, the apparatus's dynamical equation in Eq.~\eqref{generic Langevin} describes a damped oscillator under a randomly driven force. When the damping rate is larger than a critical value $\gamma>\Omega_{\rm rot}/2$, the system decays with no oscillation, known as overdamped. In the overdamped region, as the damping rate increases, the system decays to the equilibrium faster, so the random force term affects the system less. Finally, the dephasing parameter $\Gamma$ decreases as the damping rate $\gamma$ increases.

\begin{figure}
    \centering
    \includegraphics[scale=0.5]{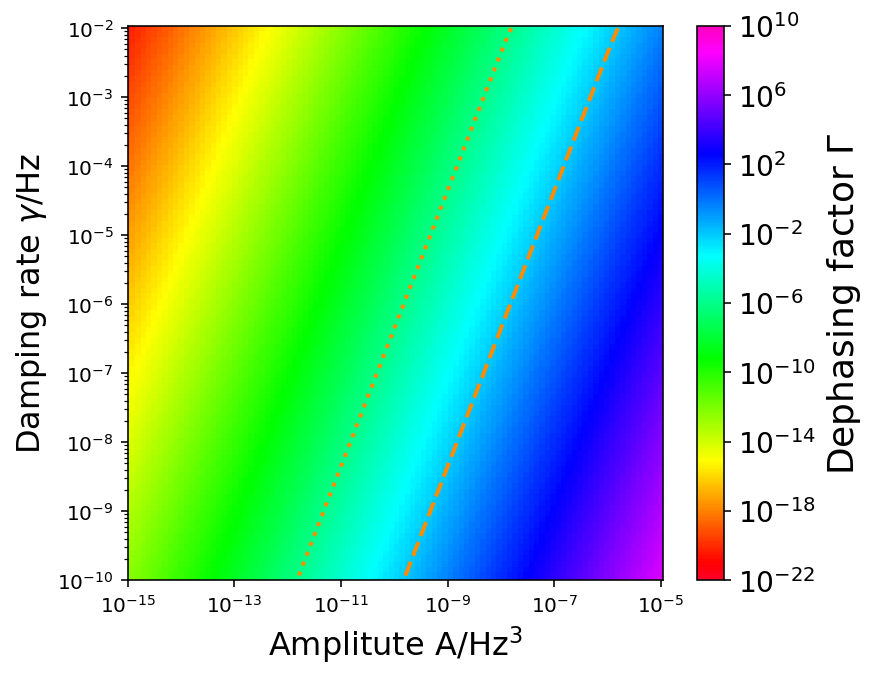}
    \caption{\small The dephasing parameter $\Gamma$ with respect to different damping rates $\gamma$ and torsion noise amplitudes $A$. The transfer function is given by Eq.~\eqref{F} with parameters $t_a=0.25$\,s and $t_e=0$\,s, and the intrinsic frequency is chosen as $\Omega_{\rm rot}=2\pi\times1$\,Hz. $\Gamma$ is required smaller than some thresholds in different situations, then the parameters should be chosen on the left side of the critical lines in the figure, where the dotted and dashed lines represent $\Gamma=10^{-6}$ and $\Gamma=0.01$ for gravimeters and QGEM experiment in respect.}
    \label{Gamma_A_gamma}
\end{figure}
Fig.~\ref{Gamma_A_gamma} shows the dependence of the dephasing parameter $\Gamma$ concerning different damping rates $\gamma$ and torsion noise amplitude $A$, where the parameters are chosen as $t_a=0.25$\,s and $t_e=0$\,s and $\Omega_{\rm rot}=2\pi\times1$\,Hz. 
As is shown, $\Gamma$ increases as $A$ increases or $\gamma$ decreases. As is discussed in section 2, the gravimeter and QGEM experiments require an upper bound of $\Gamma$ as $10^{-6}$ and $0.01$, then $A$ and $\gamma$ should be chosen in the region on the left side of the dotted and dashed critical line respectively in the figure. 

\begin{figure}
    \centering
    \includegraphics[scale=0.5]{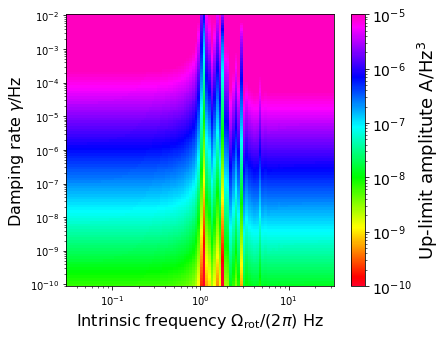}
    \caption{\small The upper bound of the amplitude $A$ assuming the value of the dephasing to be $\Gamma=0.01$. The transfer function is is given by Eq.~\eqref{F} with parameters $t_a=0.25$\,s and $t_e=0$\,s. Near the resonance region $\Omega_{\rm rot}\sim2\pi/T_{\rm tot}=2\pi\times1$\,Hz and its harmonic resonance $\Omega_{\rm rot}\sim2n\pi/T_{\rm tot}$ (with $n$ a positive integer), the upper bound of $A$ becomes more constrained. On the other hand, the requirements on $A$ relax outside the resonance region.}
    \label{Amplitute_Omega0_gamma}
\end{figure}
Fig.~\ref{Amplitute_Omega0_gamma} shows the upper bound of the torsion noise amplitude $A$ to obtain a dephasing parameter $\Gamma$ smaller than 0.01, where the parameters are also chosen as $t_a=0.25$\,s and $t_e=0$\,s. Since $\Gamma$ is proportional to $A^2$, so one can also use Fig.~\ref{Amplitute_Omega0_gamma} to analyze other threshold of $\Gamma$ by multiplying a factor on $A$. For example, one can analyze the case $\Gamma<10^{-6}$ for gravimeters by multiplying $10^{-2}$ to the upper bound of $A$.

As is shown, the restriction on the upper bound of $A$ is very stringent near the resonance region $\Omega_{\rm rot}\sim2\pi/T_{\rm tot}=2\pi\times1$\,Hz and its harmonic resonance $\Omega_{\rm rot}\sim2n\pi/T_{\rm tot}$ (with $n$ a positive integer). In particular, $A$ has to be smaller than $10^{-10}\,\rm Hz^3$ for $\gamma=10^{-10}$\,Hz, and it has to be smaller than $10^{-8}\,\rm Hz^3$ for $\gamma=10^{-2}$\,Hz. 
On the other hand, the value of $A$ is less severely constrained outside the resonance region. For a damping rate larger than $10^{-4}$\,Hz, $A$ can be larger than $10^{-5}\,\rm Hz^3$. 

In a recent simulation work~\cite{Beaufils2023}, a relationship between the torsion noise amplitude $A$ and the superposition size $\Delta x$ has been obtained as $\Delta x A\sim10^{-11}\,\rm m\cdot Hz^3$. Since the superposition size discussed in this paper is $\Delta x=a_mt_a^2=11.2\,\rm\mu m$, then the noise amplitude is $A\sim10^{-6}\,\rm Hz^3$, which is smaller than the restricted bound $A_{\rm bound}\sim10^{-5}\,\rm Hz^3$. In conclusion, as long as the damping rate is designed to be larger than $10^{-4}$\,Hz and the intrinsic frequency $\Omega_{\rm rot}$ is designed outside the resonance region, the dephasing parameter $\Gamma$ of the interferometer will be smaller than $0.01$.

\section{Summary}\label{section 6}
In this paper, we investigated inertial torsion noise (ITN) in the context of matter-wave interferometry, and some of our highlights and conclusions are summarized below.

In section \ref{section 2}, we explained the physical interpretation of the ITN and briefly reviewed the dephasing effect of generic noises on matter-wave interferometers. The key point is that the ensemble average of a random phase implies a decay factor $\mathbb{E}[\mathrm{e}^{\mathrm{i}\delta\phi}]=\mathrm{e}^{-\Gamma}$ on the off-diagonal terms of the density matrix, where the dephasing parameter is exactly the variance of the noise $\Gamma=\mathbb{E}[(\delta\phi)^2]$.

In section \ref{section 3}, we pointed out that the decay factor $\Gamma$ can be regarded as a linear response of the interferometer to the ITN. In particular, $\Gamma$ can be formulated as the PSD of the noise $S_{\dot{\Theta}^2\dot{\Theta}^2}(\omega)$ multiplying a transfer function $F(\omega)$ in frequency space. Remarkably, $F(\omega)$ only relies on the trajectories of the two arms of the interferometer and is independent of the noise. In the rest part of section 3, some asymptotic features of $F(\omega)$ are discussed, and the exact and approximate result for a certain interferometer is shown as Fig.\,\ref{transfer function}.

In section \ref{section 4}, we modelled the torsion noise by a generalized Langevin equation \eqref{generic Langevin}, which implies the PSD $S_{\Theta\Theta}(\omega)$ as \eqref{Theta PSD}. Then the PSD of the ITN $S_{\dot{\Theta}^2\dot{\Theta}^2}(\omega)$ is the self-convolution of $S_{\dot{\Theta}\dot{\Theta}}(\omega)$ according to the convolution theorem, where the math details are summarized in Appendix \ref{appendixB}. It is remarkable that the peak position of $S_{\dot{\Theta}^2\dot{\Theta}^2}(\omega)$ is doubled compared to $S_{\dot{\Theta}\dot{\Theta}}(\omega)$ because of properties of self-convolution, while the Q-factor of $S_{\dot{\Theta}^2\dot{\Theta}^2}(\omega)$ remains the same as $S_{\dot{\Theta}\dot{\Theta}}(\omega)$. 

In section \ref{section 5}, we scanned parameters theoretically and found some major features as follows. First, $\Gamma$ increases significantly near the resonance region $\Omega_{\rm rot}=2n\pi/T_{\rm tot}$. Next, $\Gamma$ descreases proportional to the damping rate $\gamma^{-1}$ in the underdamped region, while it decreases with $\gamma^{-3}$ in the overdamped region. Finally, if $\Gamma$ is required smaller than $0.01$, then constrain on the parameters $A$, $\gamma$ and $\Omega_{\rm rot}$ shown as Fig.\ref{Amplitute_Omega0_gamma}, and $A$ is tolerant up to $10^{-5}\,\rm Hz^3$ if $\gamma>10^{-4}$\,Hz and $\Omega_{\rm rot}$ is outside the resonance region.

\begin{acknowledgements}
    M. Wu would like to thank the China Scholarship Council (CSC) for financial support. MT would like to acknowledge funding from ST/W006227/1. SB thanks EPSRC grants EP/R029075/1, EP/X009467/1, and ST/W006227/1. SB and AM's research is supported by the Sloan, and Betty and Moore foundations. 
\end{acknowledgements}


\bibliographystyle{unsrt}

\appendix
\onecolumngrid

\section{Lagrangian of Inertial Torsion Noise}\label{appendixA}

In this appendix, we will derive the Lagrangian in Eq.~\eqref{Lagrangian} that gives rise to inertial torsion noise (ITN). The basic idea is firstly to construct the metric in the comoving reference frame of the experimental apparatus (Sec.~\ref{fnc}), and then to compute the Lagrangian of the test mass in the non-relativistic limit (Sec.~\ref{lagrangian}). 

\subsection{Rotating Fermi Normal Coordinates} \label{fnc}

To construct the coordinate system near the experiment, one can choose the worldline of the center of the experimental box as a fiducial time-like curve in the spacetime manifold. Based on this worldline, one can  construct the Fermi normal coordinate (FNC) of the spacetime using the method of the Fermi-Walker transport. Under this coordinate, the metric can be generally written as~\cite{Poisson2011}

\begin{equation}\label{FNC}
    \mathrm{d}s^2 = -\left[\left(1+\frac{a^bx'_b}{c^2}\right)^2+R_{0c0d}x^{\prime c}x^{\prime d}\right]c^2\mathrm{d}t^{\prime 2} 
    -\frac{2}{3}R_{0cbd}x^{\prime c}x^{\prime d}c\mathrm{d}t^{\prime}\mathrm{d}x^b
    +\left(\delta_{bc}-\frac{1}{3}R_{bdce}x^{\prime d}x^{\prime e}\right)\mathrm{d}x^{\prime b}\mathrm{d}x^{\prime c},
\end{equation}
where the indices $a,b,c,d=1,2,3$ represent the spacial coordinates. In the following, we will neglect the linear acceleration terms $\sim a^b$ and the Riemann tensor terms $\sim R_{abcd}$ such that the Metric in Eq.~\eqref{FNC} reduces to the Minkowski spacetime metric in Cartesian coordinates.

To obtain the metric in a rotating reference frame we have to make an additional transformation. Rotations along the z-axis can be described by the time-dependent angle $\theta(t)$, such that the coordinates in the rotating frame are described by

\begin{equation}\label{rotation transformation}
\begin{cases}
    t=t',\\
    x = x'\cos\theta(t)+y'\sin\theta(t), \\
    y = -x'\sin\theta(t)+y'\cos\theta(t),\\
    z= z',
\end{cases}
\end{equation}
where we recall the primed symbols represent the coordinates in the inertial (non-rotating) frame (see Fig.\,\ref{fig7}). 

To compute the metric in the rotating coordinates one can proceed is several ways. For example, one way to simplify the calculation is to write the transformation between $(x,y)$ and $(x',y')$, as well as between  $(\mathrm{d}x,\mathrm{d}y)$ and $(\mathrm{d}x',\mathrm{d}y')$, in matrix form. But here, we will offer an alternative method by exploiting the complex notation. The basic idea is to introduce a complex coordinate $w=x+\mathrm{i}y$, then the complex conjugate is $\bar{w}=x-\mathrm{i}y$, and the differentials are
\begin{equation}
\mathrm{d}w=\mathrm{d}x+\mathrm{i}\mathrm{d}y,\quad\quad \mathrm{d}\bar{w}=\mathrm{d}x-\mathrm{i}\mathrm{d}y,     
\end{equation}
with analogous expressions for the primed variables. This is a common trick to deal with 2D problems, because one can always construct a complex structure on a 2D surface through its metric to become a Riemann surface. For pedagogical material, see chapter 7 of \cite{chen1999lectures}.

The rotation transformation from Eq.~\eqref{rotation transformation}, as well as its inverse, can be simply written as
\begin{equation}
    w = w'\mathrm{e}^{\mathrm{i}\theta(t)},\quad\quad w' = w\mathrm{e}^{-\mathrm{i}\theta(t)},
\end{equation}
where we have omitted the $t$ and $z$ coordinate transformation for brevity. We find that the differential forms are

\begin{equation}
\begin{aligned}
    \mathrm{d}w^{\prime} &= \mathrm{e}^{-\mathrm{i}\theta}\mathrm{d}w' - \mathrm{i}\dot{\theta}w'\mathrm{e}^{-\mathrm{i}\theta}\mathrm{d}t, \\
    \mathrm{d}\bar{w}^{\prime} &= \mathrm{e}^{\mathrm{i}\theta}\mathrm{d}\bar{w}' + \mathrm{i}\dot{\theta}\bar{w}'\mathrm{e}^{\mathrm{i}\theta}\mathrm{d}t.
\end{aligned}    
\end{equation}
The considered terms $\mathrm{d}x^{\prime 2}+\mathrm{d}y^{\prime2}=\mathrm{d}w^{\prime}\mathrm{d}\bar{w}^{\prime}$ in the original Fermi normal coordinates from Eq.~\eqref{FNC} transform to 
\begin{equation}
\begin{split}
    \mathrm{d}w^{\prime}\mathrm{d}\bar{w}^{\prime} &= (\mathrm{e}^{-\mathrm{i}\theta}\mathrm{d}w - \mathrm{i}\dot{\theta}w\mathrm{e}^{-\mathrm{i}\theta}\mathrm{d}t)
    (\mathrm{e}^{\mathrm{i}\theta}\mathrm{d}\bar{w} + \mathrm{i}\dot{\theta}\bar{w}\mathrm{e}^{\mathrm{i}\theta}\mathrm{d}t) \\
        &= \mathrm{d}w\mathrm{d}\bar{w} - \mathrm{i}\dot{\theta}(w\mathrm{d}\bar{w}-\bar{w}\mathrm{d}w)\mathrm{d}t + \dot{\theta}^2w\bar{w}\mathrm{d}t^2 \\
        &= \mathrm{d}x^2+\mathrm{d}y^2 + 2\dot{\theta}(-x\mathrm{d}y+y\mathrm{d}x)\mathrm{d}t + \dot{\theta}^2(x^2+y^2)\mathrm{d}t^2,
\end{split}
\end{equation}
which then immediately gives the transformed metric in the rotating reference frame
\begin{equation}\label{rotating FNC}
    \mathrm{d}s^2 = -\big(c^2-\dot{\theta}^2(x^2+y^2)\big)\mathrm{d}t^2 +2\dot{\theta}(-x\mathrm{d}y+y\mathrm{d}x)\mathrm{d}t +\mathrm{d}x^2+\mathrm{d}y^2.
\end{equation}

\subsection{Lagrangian of Inertial Torsion Noise}\label{lagrangian}

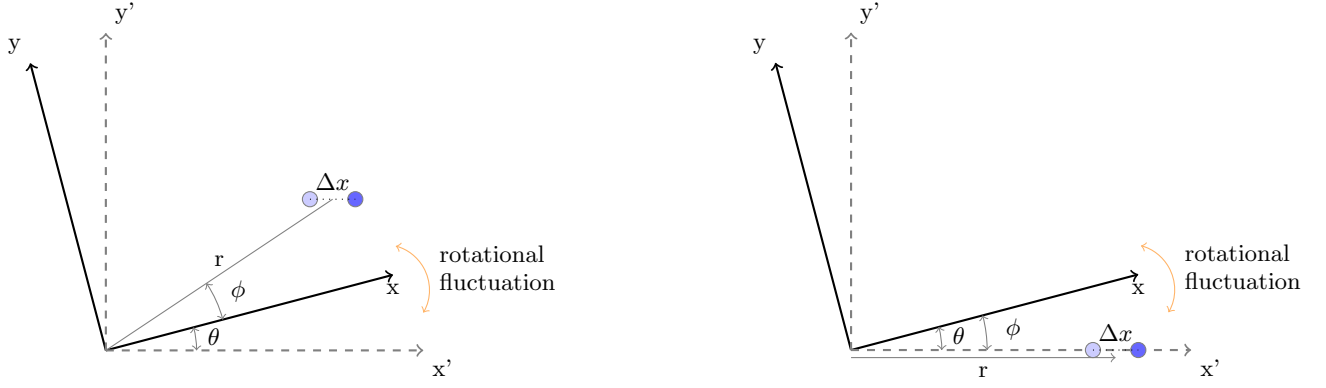
\begin{figure}
    \centering
    \begin{minipage}{0.45\textwidth}
        \begin{tikzpicture}
            \draw[->, thick] (0,0)--(3.8,1) node[below, black] {x};
            \draw[->, thick] (0,0)--(-1,3.8) node[above left, black] {y};
            \draw[->, dashed, thick, gray] (0,0)--(4.2,0) node[below right, black] {x'};
            \draw[->, dashed, thick, gray] (0,0)--(0,4.2) node[above right, black] {y'};
    
            \coordinate (x') at (4,0);
            \coordinate (x) at (3.8,1);
            \coordinate (o) at (0,0);
            \draw[<->, orange!60] (4.2,0.5) arc(-30:75:0.6);
            \node[right, align=left] at (4.3,1.1) {rotational\\ fluctuation};
            \pic["$\theta$", draw=gray, <->, angle eccentricity=1.2, angle radius=1.2cm] {angle=x'--o--x}; 
    
            \draw[gray, fill=blue!20] (2.7,2) circle (0.1);
            \draw[gray, fill=blue!60] (3.3,2) circle (0.1);
            \draw[dotted] (2.7,2)--node[above]{$\Delta x$} (3.3,2);
            \draw[gray] (0,0)--node[above, black] {r} (3,2);
            \coordinate (P) at (3,2);
            \pic["$\phi$", draw=gray, <->, angle eccentricity=1.2, angle radius=1.6cm] {angle=x--o--P};
        \end{tikzpicture}
    \end{minipage}
    \hfill
    \begin{minipage}{0.45\textwidth}
        \begin{tikzpicture}
            \draw[->, thick] (0,0)--(3.8,1) node[below, black] {x};
            \draw[->, thick] (0,0)--(-1,3.8) node[above left, black] {y};
            \draw[->, dashed, thick, gray] (0,0)--(4.5,0) node[below right, black] {x'};
            \draw[->, dashed, thick, gray] (0,0)--(0,4.2) node[above right, black] {y'};
    
            \coordinate (x') at (4,0);
            \coordinate (x) at (3.8,1);
            \coordinate (o) at (0,0);
            \draw[<->, orange!60] (4.2,0.5) arc(-30:75:0.6);
            \node[right, align=left] at (4.3,1.1) {rotational\\ fluctuation};
            \pic["$\theta$", draw=gray, <->, angle eccentricity=1.2, angle radius=1.2cm] {angle=x'--o--x}; 
    
            \draw[gray, fill=blue!20] (3.2,0) circle (0.1);
            \draw[gray, fill=blue!60] (3.8,0) circle (0.1);
            \draw[dotted] (3.2,0)--node[above]{$\Delta x$} (3.8,0);
            \draw[gray, ->] (0,-0.1)--node[below, black] {r} (3.5,-0.1);
            \pic["$\phi$", draw=gray, <->, angle eccentricity=1.2, angle radius=1.8cm] {angle=x'--o--x};
        \end{tikzpicture}
    \end{minipage}
    \caption{ Illustration of the inertial reference frame (labeled as $x'$ and $y'$) and of the rotating reference frame (labeled as $x$ and $y$), which is comoving with the experimental equipment. For simplicity, we assume that the interferometric axis (line segment connecting the light and dark blue circles) is aligned with the $x'$ axis of the inertial frame.
    (a) General case when the interferometric particle is placed at angle $\phi+\theta$ with respect to the inertial reference frame. $\phi$ denotes its polar coordinate in the rotating reference frame, and $\theta$ is the angle between the inertial and rotating reference frames. 
    (b) Special case when the interferometric axis coincides with the $x'$-axis of the inertial reference frame. In this case, the two angles $\phi$ and $\theta$ defined in point (a) have a simple relationship $\phi=-\theta$. \label{fig7}
}
\end{figure}

The Lagrangian of a point-like massive object is given by $L = -m c^2\sqrt{-\mathrm{d}s^2/(c^2\mathrm{d}t^2)}$. Using the metric in Eq.~\eqref{rotating FNC} and taking the non-relativistic limit $v_x, v_y \ll c$ (with $v_x\equiv\mathrm{d}x/\mathrm{d}t, v_y\equiv\mathrm{d}y/\mathrm{d}t$), we find that the Lagrangian is 

\begin{equation}\label{eqL}
    L = - \frac{1}{2}m\dot{\theta}^2(x^2+y^2) + m\dot{\theta}(-x v_y+v_x y),
\end{equation}
where we have omitted the constant term $mc^2$ and the kinetic energy term $1/2m(v_x^2+v_y^2)$. 
The term $-\frac{1}{2}m\dot{\theta}^2(x^2+y^2)$ describes the centrifugal force. In particular, according to Euler-Lagrange equation $\frac{\mathrm{d}}{\mathrm{d}t}\left(\frac{\partial L}{\partial\dot{x}_j}\right)-\frac{\partial L}{\partial x_j}=0$, this term gives the centrifugal force $\vec{F}_{\rm cent}=m\dot{\theta}^2\vec{r}$. 
The term $m\dot{\theta}(-xv_y+v_xy)$ will give two forces, $-2m\vec{r}\times\dot{\vec{\theta}}$ and $m\vec{r}\times{\ddot{\vec{\theta}}}$, known as the Coriolis force and the Euler force, respectively. Eq.~\eqref{eqL} is well known in the literature, and it gives rise among other things also to the Sagnac effect~\cite{post1967sagnac,anderson1994sagnac,malykin2000sagnac,torovs2020revealing}.

The Lagrangian term $m\dot{\theta}(-xv_y+v_xy)$ can be written into polar coordinates with $x=r\cos\phi$ and $y=r\sin\phi$ (see  Fig.~\ref{fig7}(a)). If $r$ is assumed constant, then $v_x=-r\dot{\phi}\sin\phi$ and $v_y=r\dot{\phi}\cos\phi$, so this term becomes $-mr^2\dot{\theta}\dot{\phi}$. In the special case shown in Fig.~\ref{fig7}(b), when the test mass is set on the $x$-axis of the inertial (non-rotating) reference frame, one may directly obtain $\phi=-\theta$. Then the Lagrangian of the Coriolis and Euler forces becomes $m\dot{\theta}^2r^2$. Therefore, the total Lagrangian of the centrifugal, the Coriolis, and the Euler forces reduces to
\begin{equation}\label{LITN}
    L_{\rm ITN} = \frac{1}{2}m\dot{\theta}^2r^2.
\end{equation}
Finally, if the angle $\theta$ is assumed to be small, then the $y$-component is much smaller than the $x$-component in the Lagrangian. Hence making the approximation $r\approx x$ we finally obtain the ITN Lagrangian in Eq.~\eqref{Lagrangian}.

\section{Calculation for PSD of ITN}\label{appendixB}

\begin{figure*}
    \centering
    \begin{minipage}{0.45\textwidth}
        \begin{tikzpicture}[scale=0.8]
            \draw[->, thick, gray] (-4,0)--(4,0) node[anchor=west, black] {Re(u)};
            \draw[->, thick, gray] (0,-2.5)--(0,4) node[anchor=south east, black] {Im(u)};
    
            \fill (1,0.15) circle (0.05) node[anchor=south] {$\omega_1$}; 
            \fill (1,-0.15) circle (0.05) node[anchor=north] {$\omega_2$};
            \fill (-1,0.15) circle (0.05) node[anchor=south east] {$-\omega_2$};
            \fill (-1,-0.15) circle (0.05) node[anchor=north east] {$-\omega_1$};
            \fill (2.5,0.15) circle (0.05) node[anchor=south west] {$\omega+\omega_1$}; 
            \fill (2.5,-0.15) circle (0.05) node[anchor=north west] {$\omega+\omega_2$}; 
            \fill (0.5,0.15) circle (0.05) node[anchor=south east] {$\omega-\omega_2$};
            \fill (0.5,-0.15) circle (0.05) node[anchor=north east] {$\omega-\omega_1$};
    
            \draw[green] (-3.75,0.05)--(3.75,0.05);
            \draw[green] (3.75,0) arc (0:180:3.75);
        \end{tikzpicture}
    \end{minipage}
    \begin{minipage}{0.45\textwidth}
        \begin{tikzpicture}[scale=0.8]
            \draw[->, thick, gray] (-4,0)--(4,0) node[anchor=west, black] {Re(u)};
            \draw[->, thick, gray] (0,-2.5)--(0,4) node[anchor=south east, black] {Im(u)};
    
            \fill (0,2) circle (0.05) node[anchor=south east] {$\omega_1$}; 
            \fill (0,1) circle (0.05) node[anchor=south east] {$\omega_2$};
            \fill (0,-2) circle (0.05) node[anchor=south east] {$-\omega_1$}; 
            \fill (0,-1) circle (0.05) node[anchor=south east] {$-\omega_2$};
            \fill (1.5,2) circle (0.05) node[anchor=south west] {$\omega+\omega_1$}; 
            \fill (1.5,1) circle (0.05) node[anchor=south west] {$\omega+\omega_2$};
            \fill (1.5,-2) circle (0.05) node[anchor=south west] {$\omega-\omega_1$}; 
            \fill (1.5,-1) circle (0.05) node[anchor=south west] {$\omega-\omega_2$};
    
            \draw[green] (-3.75,0.05)--(3.75,0.05);
            \draw[green] (3.75,0) arc (0:180:3.75);
        \end{tikzpicture}
    \end{minipage}
    \caption{ The poles and the integral path of the integral in Eq.~\eqref{SRRN definition}. The poles of $S_{\Theta\Theta}$ and $ S_{\dot{\Theta}^2\dot{\Theta}^2}(\omega)$ are given by Eq.~\eqref{Stheta poles}. Subfigure (a) and (b) show two cases where $\sqrt{\Omega_{\rm rot}^2-\gamma_{\rm rot}^2/4}$ is a real number and an imaginary number. Note that the relations $\omega_3=-\omega_2$ and $\omega_4=-\omega_1$ have been used to simplify the notation in both subfigures.}
    \label{S poles+path}
\end{figure*}
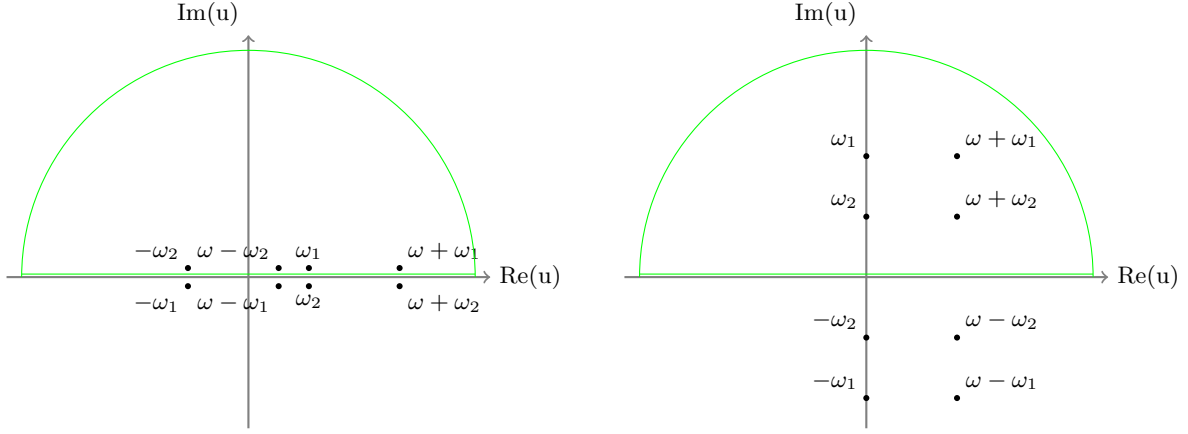

In this appendix, we will calculate the PSD of the ITN arising from a thermal environment modelled by Eq.~\eqref{rotation Langevin}. As is discussed in the main text, the PSD of ITN $S_{\dot{\Theta}^2\dot{\Theta}^2}(\omega)$ is the self-convolution of the PSD of the torsion angle $S_{\dot{\Theta}\dot{\Theta}}(\omega)=\omega^2S_{\Theta\Theta}(\omega)$, that is, 
\begin{equation}\label{SRRN definition}
\begin{split}
    S_{\dot{\Theta}^2\dot{\Theta}^2}(\omega) &= (S_{\dot{\Theta}\dot{\Theta}} \ast S_{\dot{\Theta}\dot{\Theta}})(\omega) \\
        &= \int_{-\infty}^{\infty} (2\gamma_{\rm rot}k_BT/I)^2 \frac{u^2}{(u^2-\Omega_{\rm rot}^2)^2+\gamma_{\rm rot}^2u^2} \times \frac{(\omega-u)^2}{((\omega-u)^2-\Omega_{\rm rot}^2)^2+\gamma_{\rm rot}^2(\omega-u)^2}\mathrm{d}u.
\end{split}
\end{equation}
We will use the residue theorem to calculate this integral~\cite{arfken1972mathematical}. Firstly, the poles of $S_{\Theta\Theta}(\omega)$ are
\begin{equation}\label{Stheta poles}
    \omega_{1,2,3,4} = \pm\sqrt{\Omega_{\rm rot}^2-\gamma_{\rm rot}^2/4}\pm\mathrm{i}\gamma_{\rm rot}/2.
\end{equation}
The positiveness of the discriminant $\Omega_{\rm rot}^2-\gamma_{\rm rot}^2/4$ will affect the positions of poles of the integrand in Eq.~\eqref{SRRN definition}, shown in Fig.~\ref{S poles+path}, of which the discriminants in sub-figures (a) and (b) are positive and negative respectively. Note that the relations $\omega_3=-\omega_2$ and $\omega_4=-\omega_1$ are used to simplify the notations in both sub-figures. However, since the integral is real-valued, both cases should have the same result. Thus, it is enough to consider the case $\Omega_{\rm rot}^2-\gamma_{\rm rot}^2/4\textgreater0$. 

Then according to the residue theorem, the integral value in Eq.~\eqref{SRRN definition} equals the residue value of the integrand at the poles in the path shown in Fig.~\ref{S poles+path}. In particular, this integral equals the summation of the residues at $\omega_1$, $-\omega_2$, $\omega+\omega_1$ and $\omega-\omega_2$ when $\omega\neq\omega_1+\omega_2$. In this case, every pole is a first-order pole. For the special case $\omega=\omega_1+\omega_2$, there are only two second-order poles $\omega_1$ and $\omega_2$, so the integral in Eq.~\eqref{SRRN definition} is given by these two poles. 
Since our purpose is to calculate the pure real-valued integral \eqref{SRRN definition} and different cases of poles have to give the same result, we may focus on the case $\omega\neq\omega_1+\omega_2=2\sqrt{\Omega_{\rm rot}^2-\gamma_{\rm rot}^2/4}$, then the integrand can be written as

\begin{equation}
\begin{split}
    F(u,\omega) = A^2\frac{u^2(\omega-u)^2}{(u-\omega_1)(u-\omega_2)(u+\omega_1)(u+\omega_2)} \times\frac{1}{(u-(\omega-\omega_1))(u-(\omega-\omega_2))(u-(\omega+\omega_1))(u-(\omega+\omega_2))},
\end{split}
\end{equation}

where we denote $A=2\gamma_{\rm rot}k_BT/I$ for ease of writing. Then the residue values are given by

\begin{equation}
\begin{aligned}
    2\pi\mathrm{i}\underset{u=\omega_1}{\text{Res}}F(u) &= A^2\frac{\pi\omega_1(\omega-\omega_1)^2}{2\gamma_{\rm rot}\sqrt{\Omega_{\rm rot}^2-\gamma_{\rm rot}^2/4}} \times\frac{1}{\omega(\omega-\mathrm{i}\gamma_{\rm rot})(\omega-2\sqrt{\Omega_{\rm rot}^2-\gamma_{\rm rot}^2/4})(\omega-2\omega_1)}, \\
    2\pi\mathrm{i}\underset{u=-\omega_2}{\text{Res}}F(u) &= A^2\frac{\pi\omega_2(\omega+\omega_2)^2}{2\gamma_{\rm rot}\sqrt{\Omega_{\rm rot}^2-\gamma_{\rm rot}^2/4}} \times\frac{1}{\omega(\omega-\mathrm{i}\gamma_{\rm rot})(\omega+2\sqrt{\Omega_{\rm rot}^2-\gamma_{\rm rot}^2/4})(\omega+2\omega_2)}, \\
    2\pi\mathrm{i}\underset{u=\omega+\omega_1}{\text{Res}}F(u) &= A^2\frac{(\omega+\omega_1)^2\omega_1^2}{\omega(\omega+\mathrm{i}\gamma_{\rm rot})(\omega+2\omega_1)(\omega+2\sqrt{\Omega_{\rm rot}^2-\gamma_{\rm rot}^2/4})} \times\frac{\pi}{2\omega_1\gamma_{\rm rot}\sqrt{\Omega_{\rm rot}^2-\gamma_{\rm rot}^2/4}}, \\
    2\pi\mathrm{i}\underset{u=\omega-\omega_2}{\text{Res}}F(u) &= A^2\frac{(\omega-\omega_2)^2\omega_2^2}{\omega(\omega-2\sqrt{\Omega_{\rm rot}^2-\gamma_{\rm rot}^2/4})(\omega-2\omega_2)(\omega+\mathrm{i}\gamma_{\rm rot})} \times\frac{\pi}{2\omega_2\gamma_{\rm rot}\sqrt{\Omega_{\rm rot}^2-\gamma_{\rm rot}^2/4}}.
\end{aligned}
\end{equation}

Finally the PSD of ITN defined as the integral \eqref{SRRN definition} is 

\begin{equation}
\begin{split}
    S_{\dot{\Theta}^2\dot{\Theta}^2}(\omega) &= 2\pi\mathrm{i}(\underset{u=\omega_1}{\text{Res}}F(u)+\underset{u=-\omega_2}{\text{Res}}F(u)+\underset{u=\omega+\omega_1}{\text{Res}}F(u)+\underset{u=\omega-\omega_2}{\text{Res}}F(u)) \\
        &= A^2\frac{\pi}{\gamma_{\rm rot}} \frac{4\omega^4+4(\gamma_{\rm rot}^2-3\Omega_{\rm rot}^2)\omega^2+16\Omega_{\rm rot}^4}{(\omega^2+\gamma_{\rm rot}^2)(4\gamma_{\rm rot}^2\omega^2+(\omega^2-4\Omega_{\rm rot}^2)^2)}.
\end{split}
\end{equation}

\section{Gas collision noise}\label{appendixC}

In this appendix, we will consider a certain source of the inertial torsion noise, that is, the collision due to thermal motion of gas molecules. According to the fluctuation-dissipation theorem, the applitute of the random force due to the gas collision is $\sqrt{A}=\sqrt{2\gamma_{\rm rot}k_BT/I}$, where $k_B$ is the Boltzmann constant and $T=300$\,K is the gas temperature outside the experiment box. Then the dynamical equation \eqref{generic Langevin} of the torsion motion of the experiment box becomes to the Langevin equation~\cite{lemons1997paul}
\begin{equation}\label{rotation Langevin}
    \ddot{\Theta} = -\Omega_{\rm rot}^2\Theta - \gamma_{\rm rot}\dot{\Theta} + \sqrt{2\gamma_{\rm rot}k_BT/I}\Theta_{\rm in},
\end{equation}
In this case, the power spectrum for $\Theta$ and $\dot{\Theta}^2$ are
\begin{equation}
\begin{aligned}
    S_{\Theta\Theta}(\omega) &= \frac{2\gamma_{\rm rot}k_BT/I}{(\Omega_{\rm rot}^2-\omega^2)^2+\gamma_{\rm rot}^2\omega^2}, \\
    S_{\dot{\Theta}^2\dot{\Theta}^2}(\omega) &= \frac{4\pi\gamma_{\rm rot}(k_BT)^2}{I^2}\frac{4\omega^4+4(\gamma_{\rm rot}^2-3\Omega_{\rm rot}^2)\omega^2+16\Omega_{\rm rot}^4}{(\omega^2+\gamma_{\rm rot}^2)(4\gamma_{\rm rot}^2\omega^2+(\omega^2-4\Omega_{\rm rot}^2)^2)}.
\end{aligned}
\end{equation}
It is notatble that $S_{\dot{\Theta}^2\dot{\Theta}^2}(\omega)$ has an additional $\gamma^2$ dependence in comparison to the general case in Eq.~\eqref{ITN PSD}. This difference arises because the amplitude $A=2\gamma_{\rm rot}k_BT/I$ of the external force for the thermal noise in Eq.~\eqref{rotation Langevin} is proportional to the damping rate $\gamma_{\rm rot}$, which does not hold for the general case considered in this section. 

Note that in this case, the damping rate $\gamma_{\rm rot}$ describes both the dissipation effect and the random force caused by the collision from the ambient thermal gas molecules, which is given by \cite{PhysRevLett.103.140601, CAVALLERI20103365}
\begin{equation}\label{gamma_rot}
    \gamma_{\rm rot} = \frac{L^4}{I}\left(1+\frac{\pi}{12}\right)P_{\rm gas}\sqrt{\frac{2m_{\rm gas}}{\pi k_BT}},
\end{equation}
where $P_{\rm gas}$ is the pressure of gas and $m_{\rm gas}$ is the mass of the gas molecules. Since $\gamma_{\rm rot}$ is proportional to $P_{\rm gas}$, it can vary depending on the gas pressure outside the box. Apart from $P_{\rm gas}$, all other factors contribute a factor around $10^{-4}\sim10^{-3}$\,Hz/Pa, so the value of the damping rate $\gamma_{\rm rot}$ can be estimated as
\begin{equation}
    \gamma_{\rm rot}/[\rm Hz] \sim 10^{-4}P_{\rm gas}/[\rm Pa].
\end{equation}
For instance, in the atmosphere pressure $P_{\rm gas}=10^5$\,Pa, the damping rate is $\gamma_{\rm rot}\sim10^1$\,Hz. When the gas pressure outside the experiment box is pumped as $10^2$\,Pa or $10^{-6}$\,Pa by a rough-vacuum pump or a series of ultra-high-vacuum pumps respectively, the corresponding damping rates $\gamma_{\rm rot}$ are $10^{-2}$\,Hz and $10^{-10}$\,Hz for respect. Note that a value $\gamma_{\rm rot}\sim10^{-9}$\,Hz has already been measured in experiment~\cite{PhysRevLett.103.140601}.
\begin{figure}
    \centering
    \includegraphics[scale=0.5]{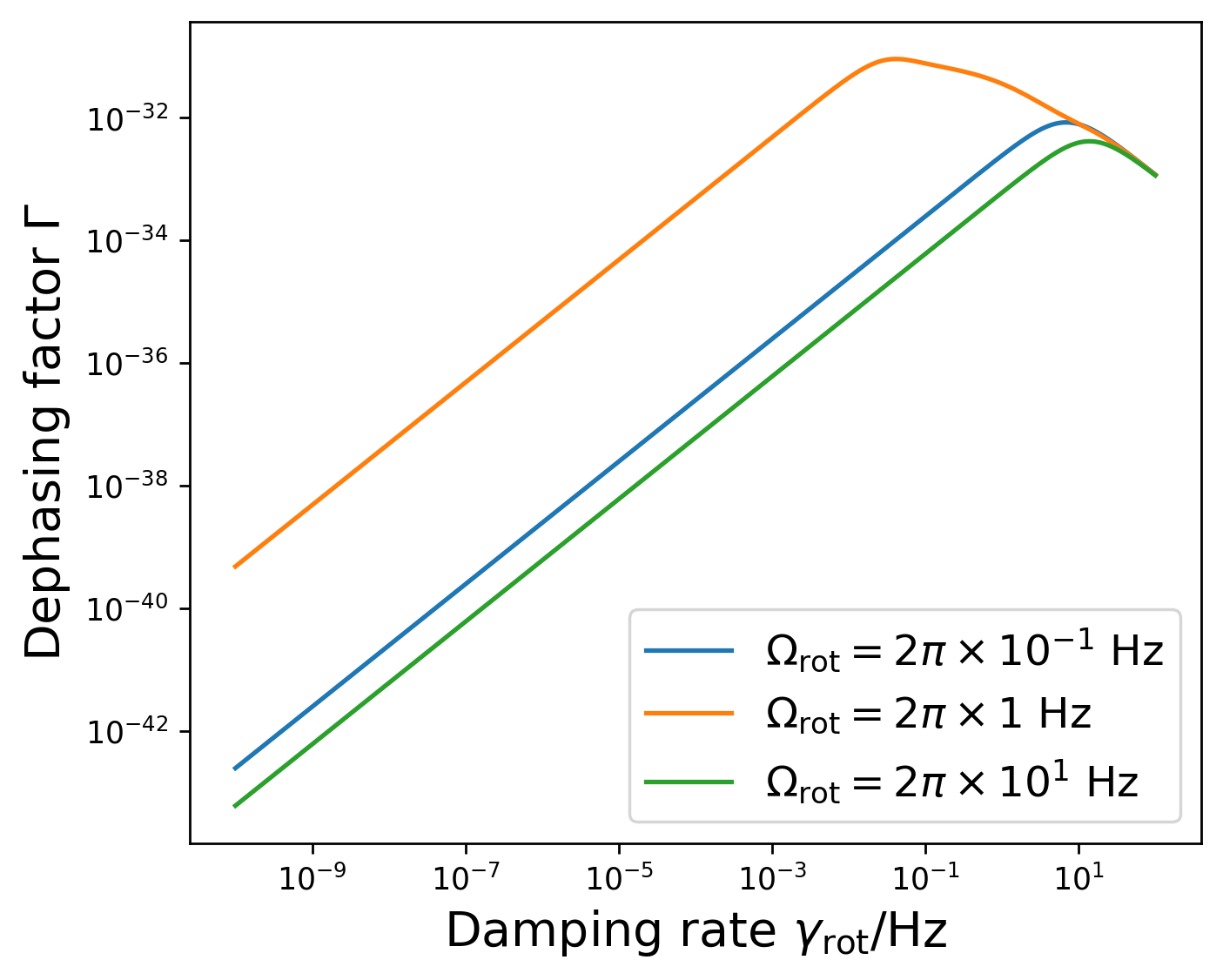}
 \caption{Dephasing factor $\Gamma$ as a function of the damping factor $\gamma_{\rm rot}$ for the thermal case. The transfer function is given in Eq.~\eqref{F} with parameters $t_a=0.25$\,s and $t_e=0$\,s. Three different $\Omega_{\rm rot}$ are chosen in the low rotation frequency limit (blue line), the resonance frequency condition $\Omega_{\rm rot}\sim2\pi/T=2\pi$\,Hz (orange line), and the high rotation frequency limit (green line). The dephasing $\Gamma$ is proportional to $\gamma_{\rm rot}$ when $\gamma_{\rm rot}\ll\Omega_{\rm rot}$, while it starts decreasing with respect to $\gamma_{\rm rot}$ when $\gamma_{\rm rot}$ becomes comparable or larger than $\Omega_{\rm rot}$.}
    \label{Gamma-gamma}
\end{figure}

As for the dephasing factor, Fig.~\ref{Gamma-gamma} shows the dependence of $\Gamma$ on the damping factor $\gamma_{\rm rot}$ caused by the thermal gas molecules. As is shown, for the thermal case, $\Gamma$ is proportional to $\gamma_{\rm rot}$ and $\gamma_{\rm rot}^{-1}$ in the underdamped region and the overdamped region in respect, which has an additional $\gamma^2$ dependence in comparison to the general case because the thermal amplitude $A^2$ is proportional to $\gamma_{\rm rot}^2$.

A final remark on Fig.\ref{Gamma-gamma} is that the dephasing factor $\Gamma$ caused by the torsion noise from gas molecules collision is negligibly tiny. Two main reasons cause this. First, the thermal motion of gas molecules is proportional to a small thermal factor $k_BT\sim10^{-21}$\,J. Besides, ITN is a second-order effect, so every little factor in the PSD $S_{\Theta\Theta}(\omega)$ of the experiment apparatus will be squared for the dephasing factor $\Gamma$. Combining both effects, the dephasing factor $\Gamma$ is suppressed by a highly tiny factor $(k_BT)^2\sim10^{-42}\,\rm J^2$, such that $\Gamma$ does not exceed $10^{-30}$.

\end{document}